\documentclass[aps,prx,floatfix,nopacs,superscriptaddress,twocolumn,reprint]{revtex4-2}

\usepackage[utf8]{inputenc}
\usepackage[american,]{babel}
\usepackage[T1]{fontenc}
\usepackage[pdftex]{graphicx}  
\usepackage{graphicx, xcolor}
\usepackage{amsmath,amsthm,amssymb}
\usepackage{hyperref}
\usepackage{xcolor}
\hypersetup{colorlinks,bookmarksopen,bookmarksnumbered,
citecolor=[rgb]{0.255,0.412,0.882},
linkcolor=[rgb]{0.255,0.412,0.882},
pdfstartview=false,
urlcolor=[rgb]{0.255,0.412,0.882}}

\renewcommand{\Im}{\operatorname{Im}}

\begin{document}

\title{Diffusion and Thermalization in a Boundary-Driven Dephasing Model}

\author{Xhek Turkeshi}
\email{xturkesh@sissa.it}
\affiliation{SISSA, via Bonomea 265, 34136 Trieste, Italy}
\affiliation{INFN, via Bonomea 265, 34136 Trieste, Italy}
\affiliation{The Abdus Salam International Centre for Theoretical Physics, strada Costiera 11, 34151 Trieste, Italy}
\author{M. Schir\'o}\thanks{ On Leave from: Institut de Physique Th\'{e}orique, Universit\'{e} Paris Saclay, CNRS, CEA, F-91191 Gif-sur-Yvette, France}
\affiliation{JEIP, USR 3573 CNRS, Coll\'ege de France,   PSL  Research  University, 11,  place  Marcelin  Berthelot,75231 Paris Cedex 05, France}

\begin{abstract}
We study a model of non-interacting spinless fermions coupled to local dephasing and boundary drive and described within a Lindblad master equation. The model features an interplay between infinite temperature thermalization due to bulk dephasing and a non-equilibrium stationary state due to the boundary drive and dissipation. We revisit the linear and non-linear transport properties of the model, featuring a crossover from ballistic to diffusive scaling, and compute the spectral and occupation properties encoded in the single particle Green's functions, that we compute exactly using the Lindblad equations of motion in spite of the \emph{interacting} nature of the dephasing term. 
We show that the distribution function in the bulk of the system becomes frequency independent and flat, consistent with infinite temperature thermalization, while near the boundaries it retains strong non-equilibrium features that reflect the continuous injection and depletion of particles due to driving and dissipation.
\end{abstract}

\date\today

\maketitle

\section{Introduction}
\label{sec:intro}

Transport and thermalization are two fundamental phenomena that occur at the macroscopic scale. Their microscopic understanding has been the subject of a long standing effort both in the classical statistical mechanics context~\cite{bertini2015macroscopic}, as well as in the quantum domain~\cite{bertini2020finitetemperature}.
One can generically distinguish two settings, depending on whether the system remains isolated from, or interacts with an environment. 
In the former setup, many-body interactions play the key role in establishing the time scales for transport of conserved charges and thermalization, whereas in the latter the environment dissipation is responsible for the late time stationary state. 

A specific instance that has attracted interest is the one of open Markovian systems~\cite{breuerPetruccione2010}, in which the system is intrinsically mixed and described by a master equation for the density matrix. Here the competition between unitary and dissipative couplings can lead to non-equilibrium phase transitions between phases with different transport properties.

In this context, finding a simple model which captures the key physical aspects of transport and thermalization while being computationally tractable for large system sizes is particularly important.
An example that has attracted specific interest is the so called dephasing model, describing a free fermionic lattice system coupled to a local Markovian environment described by a jump operator proportional to the density of particles. This model has been introduced in Ref.~\cite{esposito2005exactly,esposito2005emergence} as a toy model for the emergence of diffusive behavior for a single-particle system. The spectral properties of the model in terms of the Liouvillian spectrum have been extensively discussed~\cite{eisler2011crossover} as well as its transport properties~\cite{znidaric2010exact,znidaric2011transport,znidaric2010a,znidaric2011solvable,znidaric2013transport,znidaric2016diffusive} when the model is supplemented by a boundary Markovian driving. The pure dephasing model is furthermore exactly solvable ~\cite{medvedyeva2016exact,ziolkowska2020yangbaxter}, its Lindbladian spectrum being obtainable through Bethe Ansatz techniques~\cite{ziolkowska2020yangbaxter,essler2020integrability,buca2020dissipative}. 
Recent works have discussed various extensions of this model~\cite{bastianello2020generalized}.

In this work we revisit the transport properties of the dephasing model with boundary drive, and connect them to the onset of thermalization as defined from the emergence of a fluctuation-dissipation theorem in the single particle Green's functions that we show to be exactly computable. We focus in particular on the crossover/transition as the system size is increased between infinite temperature thermalization due to the bulk dephasing and the onset of a non-equilibrium steady state current at finite size due to the boundary drive. We investigate the role of the dephasing rate on this scenario, as well as of the deviation from the linear response regime as the strength of the drive is increased.

This paper is structured as follows. In Sec.~\ref{sec:model} we introduce the model and the non-equilibrium protocol. In Sec.~\ref{sec:obs} we discuss the main quantities of interest, namely the stationary current and the local Green's functions, and summarize the results on how to compute them within the Lindblad equation. In Sec.~\ref{sec:results} we present numerical results, first recalling the transport properties of the model and then discussing thermalization and effective temperature. We detail in the Appendix derivations of the various results used in the main text.

\begin{figure}[h!]
   \includegraphics[width=\columnwidth]{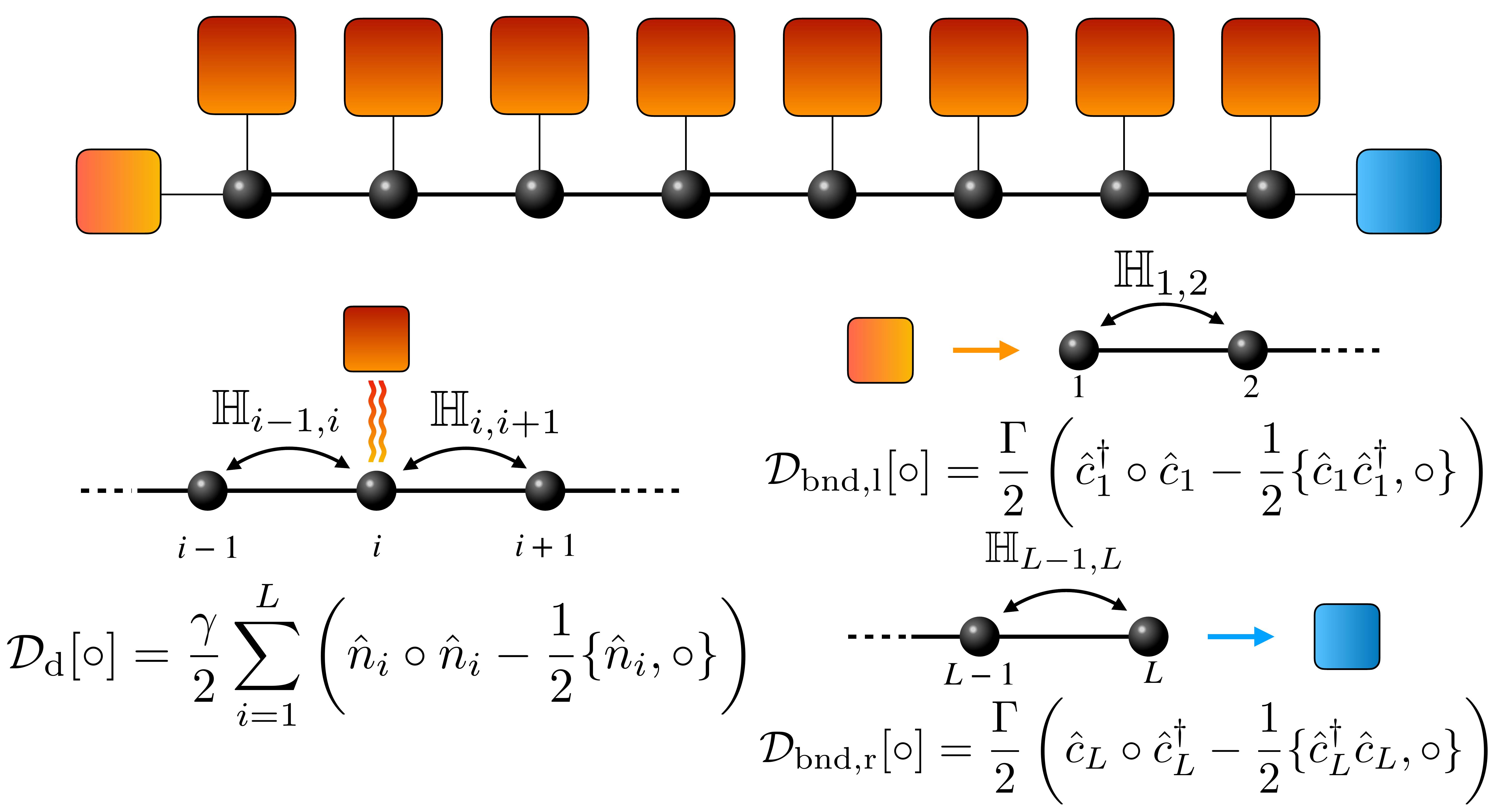}
   \caption{\label{fig:boundary_drive} Cartoon summarizing the system when the boundary driving is open. The boundary terms inject/absorb particles from the system. In addition, there is a uniform dephasing rate $\gamma$ on each lattice site.}
\end{figure}

\section{Model}
\label{sec:model}
We consider an open Markovian quantum system whose density matrix evolves according to the Lindblad equation
\begin{align}
	\partial_t \hat{\rho}_t &= \mathcal{L}\hat{\rho}_t = -i [\hat{H},\hat{\rho}_t] + \mathcal{D}[\hat{\rho}_t],\label{eq:lindblad}
\end{align}
We denote with $[\hat{A},\hat{B}]$ ( $\{\hat{A},\hat{B}\}$) the commutator (anticommutator) between $\hat{A}$ and $\hat{B}$. In Eq.~\eqref{eq:lindblad}, $\mathcal{H}$ is the Hamiltonian of the system, while the dissipation superoperator $\mathcal{D}$ collects the dissipation/dephasing terms. We set these object in the following.  

We consider a system of non-interacting spinless fermions defined in a one-dimensional chain of $L$ sites with next-neighboring hopping rate $J$
\begin{align}
	\hat{H}&= \sum_{i=1}^{L-1} J (\hat{c}^\dagger_i \hat{c}_{i+1} + \text{h.c.}) + h \hat{c}^\dagger_i \hat{c}_{i}  = \sum_{i,j=1}^N \hat{c}^\dagger_i \mathbb{H}_{i,j}\hat{c}_j,\nonumber\\
 \mathbb{H}_{i,j} & = J\delta_{i,j+1} + J\delta_{i+1,j} + h \delta_{i,j}.\label{eq:H}
\end{align}
In Eq.~\eqref{eq:H} $\hat{c}_i$ ($\hat{c}^\dagger_i$) are the annihilation (creation) fermionic operators acting on the $i$-th site, the one-particle Hamiltonian $\mathbb{H}$ is a $L\times L$ hermitian and tridiagonal matrix, and we consider open boundary conditions. 

We couple the left and right edge sites (Fig.~\ref{fig:boundary_drive}) respectively to a source and a sink. Specifically, particles are injected with a rate $\propto \Gamma$ on the first site, and are depleted with the same rate on the $L$-th site.
These two terms are responsible for a direct current (DC) in the stationary state. In addition, on every lattice site we include a dephasing bath with rate $\propto \gamma$, which provide an energy relaxation channel. These three processes are collectively captured by the following dissipators
\begin{align}
	\displaystyle\mathcal{D}[\circ]&= \mathcal{D}_\mathrm{d}[\circ] + \mathcal{D}_\mathrm{bnd,l}[\circ] + \mathcal{D}_\mathrm{bnd,r}[\circ],\label{eq:ffA1}\\
	\mathcal{D}_\mathrm{d}[\circ] &= \frac{\gamma}{2}\sum_{i=1}^L \left(\hat{n}_i\circ  \hat{n}_i - \frac{1}{2}\left\lbrace \hat{n}_i,\circ\right\rbrace\right) \label{eq:ffA1b}\\
	\mathcal{D}_\mathrm{bnd,l}[\circ] &= \frac{\Gamma}{2} \left(\hat{c}^\dagger_1\circ \hat{c}_1 - \frac{1}{2}\left\lbrace \hat{c}_1\hat{c}^\dagger_1,\circ\right\rbrace\right) \label{eq:ffA1c}\\
	\mathcal{D}_\mathrm{bnd,r}[\circ] &= \frac{\Gamma}{2} \left(\hat{c}_L\circ \hat{c}^\dagger_L - \frac{1}{2}\left\lbrace \hat{c}^\dagger_L\hat{c}_L,\circ\right\rbrace\right).\label{eq:ffA1d}
\end{align}
Above, we introduced the fermion number $\hat{n}_i \equiv \hat{c}^\dagger_i \hat{c}_i$. Eq.~\eqref{eq:ffA1b} is the dephasing contribution, Eq.~\eqref{eq:ffA1c} is the particle-injection contribution acting on the first site, while Eq.~\eqref{eq:ffA1d} is the particle-ejection contribution acting on the last site.

We emphasize that this model is not quadratic due to the dephasing processes, yet as we will discuss, it is simpler than a fully interacting many body system while retaining some interesting non-trivial aspects. 
The boundary driving introduces a further source of translational invariance breaking (other than the open boundary condition), which manifests in the presence of a stationary current. (See however Appendix~\ref{app:trans} for the translational invariant version of the model discussed in this paper). 

In the following, we are interested in characterizing the stationary state of the system obtained under the Markovian evolution~\eqref{eq:lindblad}. In particular, we will focus on the single particle correlation functions at equal times, and on the single particle Green's functions, from which physical quantities can be obtained such as the average current or the local density of states.
In the next section, we introduce these quantities and discuss how to evaluate them exactly for our model.

\section{Observables}
\label{sec:obs}

\subsection{Equal-time correlation function and current}
\label{subsec:equal}
We introduce the single-particle correlation matrix $C$, with elements
\begin{align}
	{C}_{n,m}(t) = \langle \hat{c}_n \hat{c}^\dagger_m\rangle_t  =\mathrm{tr}(\rho(t) \hat{c}_n \hat{c}^\dagger_m).
	\label{eq:defcorr}
\end{align}
This function encode the average particle density
\begin{equation}
	n_m(t) \equiv \langle \hat{n}_m\rangle_t = \mathrm{tr}(\rho(t)\hat{c}^\dagger_m\hat{c}_m) =1-C_{m,m}(t),\label{eq:pardens}
\end{equation}
as well as the average current operator. We note in fact that both the Hamiltonian and the dephasing dissipator commute with the total particle number, and therefore allows to define a conserved current which is then transported due to the boundary drive terms. This reads, from the continuity equation for the Hamiltonian part only (cfr. Eq.~\eqref{eq:H})
\begin{align}
	\hat{j}_m \equiv  i J(\hat{c}^\dagger_{m-1} \hat{c}_m - \hat{c}^\dagger_m \hat{c}_{m-1}) \nonumber \\
	j_m(t) \equiv \langle \hat{j}_m\rangle_t = 2 J \Im{C}_{m,m-1}(t).\label{eq:current}
\end{align}

Its dynamics can be obtained by the Lindblad equation Eq.~\eqref{eq:lindblad}, with dissipator Eq.~\eqref{eq:ffA1}, and using the cyclic property of the trace~\cite{breuerPetruccione2010}. We have
\begin{align}
\begin{split}
	\partial_t {C}_{m,n}(t) &= -i J {C}_{m-1,n}(t) - i J {C}_{m+1,n}(t)  +
	\\
	 & + i J {C}_{m,n+1}(t)+ i J {C}_{m,n-1}(t)  +
	 \\
	 & +\Gamma \delta_{m,N}\delta_{n,N} - \gamma (1-\delta_{m,n}) C_{m,n}(t) +
	 \\
	 &   -\frac{\Gamma}{2} (\delta_{1,n} + \delta_{1,m} + \delta_{N,n} + \delta_{N,m}) C_{m,n}(t).
	\label{eq:eom_boundary}
\end{split}
\end{align}
Notice that here we imposed $C_{0,k}(t)=C_{k,N+1}(t) = 0$ at every timestep in Eq.~\eqref{eq:eom_boundary}, as open boundary conditions apply.

Remarkably, despite the dephasing acts as a four body interaction, the equation of motion for the two-point function is closed and can be exactly computed. 
In fact, for the system of interest, the $k$-point equal time correlation function presents a hierarchical structure, where the $k$-point equal time correlation function decouple from higher order ones~\cite{eisler2011crossover}, whereas the state is non-Gaussian~\cite{dolgirev2020nongaussian} and highly entangled~\cite{alba2021spreading}.

It is convenient to rephrase Eq.~\eqref{eq:eom_boundary} in matrix form. The terms proportional to the imaginary unit $i$ are due to the commutation with the Hamiltonian, which in the fermionic representation is completely specified by $\mathbb{H}$ (cfr. eq.~\eqref{eq:H}).
The remaining terms are related to the dissipation. We define the matrix $\mathbb{D}$, whose elements are $\mathbb{D}_{m,k} = \delta_{m,k}(\gamma +\delta_{m,1} \Gamma + \delta_{m,N} \Gamma)/2$, and the matrix $P$, whose elements are $P_{m,k}(t) = \delta_{m,k} (\gamma C_{m,m}(t) + \Gamma \delta_{m,N}\delta_{k,N})\equiv \delta_{m,k} p_k(t)$. 
With these choices, and defining we have
\begin{align}
	\partial_t C = -i [\mathbb{H},C] - \{ \mathbb{D},C\} + P\equiv -i\mathbb{H}_\mathrm{eff}  C + i C \mathbb{H}_\mathrm{eff}^\dagger + P.\label{eq:tmat}
\end{align}
In the last step, we defined the first-quantization non-Hermitian Hamiltonian. In fact, manipulating Eq.~\eqref{eq:lindblad} we obtain the natural non-Hermitian Hamiltonian
\begin{align}
	\hat{H}_\mathrm{eff} = \sum_{i,j=1}^L \hat{c}^\dagger_i \mathbb{H}_\mathrm{eff} \hat{c}_j.\label{eq:nonherm}
\end{align}

In the following we show that the spectral properties of the non-Hermitian Hamiltonian determine the stationary values of the particle density and current, and of the Green's function (Sec.~\ref{subsec:green}).

Despite the simulation of the whole dynamics Eq.~\eqref{eq:tmat} is polynomial in resources, the stationary solution is obtained exactly by solving $\partial_t C=0$. 
This is given by~\cite{varma2017fractality,taylor2021subdiffusion}
\begin{align}
	C(\infty) = \int_0^\infty d\tau e^{-i \mathbb{H}_\mathrm{eff}\tau} P(\infty) e^{i \mathbb{H}_\mathrm{eff}^\dagger \tau}.\label{eq:exactsolss}
\end{align}
This integral equation is simplified once the eigendecomposition of the matrix $\mathbb{H}_\mathrm{eff}$ is considered (See Appendix~\ref{app:heff} for a coordinate Bethe ansatz solution). Its non-hermitian nature gives rise to inequivalent left and right eigenvectors: $ \mathbb{H}_\mathrm{eff} = \sum_p \lambda_p |\varphi^{\mathrm{R}}(p)\rangle \langle \varphi^{\mathrm{L}}(p)|$. 
We fix the ambiguity in the eigenvector by choosing them biorthonormal $\langle \varphi^\mathrm{L}(q)|\varphi^\mathrm{R}(p)\rangle = \delta_{p,q}$. In this way, the series expansion reduces to the trivial
\begin{align}
	e^{i\tau \mathbb{H}_\mathrm{eff} } = \sum_{k=1}^L e^{i \tau \lambda_k}|\varphi^\mathrm{R}(k)\rangle \langle \varphi^\mathrm{L}(k)|.
	\label{eq:eigdectmat}
\end{align}
Defining the components $\varphi_{i}^{\mathrm{R}/\mathrm{L}}(p)$, and performing the integral, Eq.~\eqref{eq:exactsolss} transpose to
\begin{align}
	C_{i,j}(\infty) &= \Gamma \Theta_{i,j,L} + \gamma \sum_{k=1}^L \Theta_{i,j,k} C_{k,k}(\infty)\nonumber \\
	\Theta_{i,j,l} &\equiv -\sum_{p,q=1}^L \frac{\varphi^\mathrm{R}_i(p)(\varphi^\mathrm{L}_l(p))^*(\varphi^\mathrm{R}_j(q))^*\varphi^\mathrm{L}_l(q)}{i\lambda_p - i\lambda^*_q}.\label{eq:thetaf1}
\end{align}
Solving this linear system we obtain both the density of particles $n_m(\infty)$ (cfr. Eq.~\eqref{eq:pardens}) and the stationary current $j_m(\infty)$ (cfr. Eq.~\eqref{eq:current}). 

We point out that the knowledge of the full tensor $\Theta$ is not required to obtain $n_m$ and $j_m$. In fact, the density profile requires $\mathcal{O}(N^2)$ terms, whereas the stationary current only $\mathcal{O}(1)$ as $j_i(\infty) \equiv j_\infty$ is flat.

The latter is a consequence of the generalized continuity equation in the present setup. We conclude this section by showing that the stationary current is indeed constant throughout the system.

In the non-equilibrium steady state (NESS) we have
\begin{subequations}
	\begin{align}
	0&  = \frac{d}{dt} n_1 \nonumber \\
		& = i \langle [\hat{H},\hat{c}_1^\dagger \hat{c}_1\rangle_\mathrm{NESS}  + \mathrm{tr}(\hat{c}_1^\dagger \hat{c}_1 \mathcal{D}_\mathrm{bnd,l}[\rho_\infty])\label{eq:firstsite}\\
	0&  = \frac{d}{dt} n_k = i \langle [\hat{H},\hat{c}_k^\dagger \hat{c}_k\rangle_\mathrm{NESS},\quad 2\le k\le L-1\label{eq:bulksite}\\
	0 &  = \frac{d}{dt} n_L \nonumber \\
		& = i \langle [\hat{H},\hat{c}_L^\dagger \hat{c}_L\rangle_\mathrm{NESS}+  \mathrm{tr}(\hat{c}_L^\dagger \hat{c}_L \mathcal{D}_\mathrm{bnd,r}[\rho_\infty])\label{eq:lastsite}
\end{align}
\end{subequations}
All the other terms, not reported, are zero. In fact, the dephasing damps the evolution of all the off-diagonal correlation matrix elements $C_{m,n}$ ($m\neq n$), while does not contribute to the equation of motion of the particle density.

From the bulk equation, after a simple manipulation, we find
\begin{align}
	j_k = j_{k+1},\qquad &2\le k\le L-1.
\end{align}
Let us define 
\begin{align}
	j_\mathrm{l} &\equiv \Gamma (1-n_1)\\
	j_\mathrm{r} &\equiv \Gamma n_L,\label{eq:edgecurr}
\end{align}
respectively, the current incoming/outgoing into/from the system. These contribution are those arising from $\mathcal{D}_\mathrm{bnd,l/r}$ in Eq.~\eqref{eq:firstsite} and Eq.~\eqref{eq:lastsite} respectively. Hence, we find
\begin{align}
	j_\mathrm{l} = j_2,\qquad j_\mathrm{r} = j_L.
\end{align}
Since the stationary current is constant throughout the system, we define the stationary current by the value $j_\infty\equiv j_k$ for any $k$ evaluated at the steady state.

\subsection{Non-equilibrium Green's functions}
\label{subsec:green}
A richer information on the structure of the stationary state of the Lindblad master equation can be extracted from the non-equilibrium Green's functions, which describe the excitations on top of the NESS~\cite{breuerPetruccione2010,dorda2014auxiliary,scarlatella2019spectral}. We define in particular the retarded/advanced/Keldysh Green's functions $G^{R/A/K}(t)$
\begin{subequations}
\label{eq:nesskeldysh}
	\begin{align}
	G^R_{m,n}(t)& = -i \theta(t)\langle \{\hat{c}_m(t),\hat{c}^\dagger_n\}\rangle_\mathrm{NESS}, \label{eq:defrit}\\
	G^A_{m,n}(t)& = i \theta(-t)\langle \{\hat{c}_m(t),\hat{c}^\dagger_n\}\rangle_\mathrm{NESS}, \label{eq:defadv}\\
	G^K_{m,n}(t)& = -i \langle [\hat{c}_m(t),\hat{c}^\dagger_n]\rangle_\mathrm{NESS} \label{eq:defkeld},
\end{align}
\end{subequations}
and their Fourier transforms
\begin{align}
	G^{R/A/K}(\omega) = \int_{-\infty}^{\infty} dt e^{i\omega t} G^{R/A/K}(t).
\end{align}
Since the stationary state is time-translational invariant, we have
\begin{align}
	G^{R/A}(\omega) = G^{A/R\dagger}(\omega),\qquad G^K(t) = -G^{K\dagger}(-t),\label{eq:properties}
\end{align}
where the matrix notation is understood. 

In the definition above the average over the NESS is defined as
\begin{align}
	\langle X\rangle_\mathrm{NESS} \equiv \mathrm{tr}(\hat{\rho}_\infty X).
\end{align}
while we have introduced the evolution of the operator $\hat{c}_m(t)$, which is given by the modified adjoint Lindbladian
\begin{align}
	\frac{d}{dt}\hat{c}_m(t)& = \tilde{\mathcal{L}}^\dagger\hat{c}_m(t), \label{eq:modadj}
\end{align}
with $\underline{\mathcal{L}}$ defined as
\begin{align}
	&\tilde{\mathcal{L}}^\dagger(\circ) \equiv i[\hat{H},\circ] + \mathcal{D}_\mathrm{d}[\circ] + \tilde{\mathcal{D}}_\mathrm{bnd,l}[\circ] + \tilde{\mathcal{D}}_\mathrm{bnd,r}[\circ],\\
	&\tilde{\mathcal{D}}_\mathrm{bnd,l}[\circ] = \Gamma \left(\eta \hat{c}_1\circ \hat{c}^\dagger_1 - \frac{1}{2}\left\lbrace \hat{c}_1\hat{c}^\dagger_1,\circ\right\rbrace\right)\\
	&\tilde{\mathcal{D}}_\mathrm{bnd,r}[\circ] = \Gamma \left(\eta \hat{c}^\dagger_L\circ \hat{c}_L - \frac{1}{2}\left\lbrace \hat{c}^\dagger_L\hat{c}_L,\circ\right\rbrace\right).
\end{align}
In the above equation $\hat{H}$ and $\mathcal{D}_\mathrm{d}$ are as in, respectively, Eq.~\eqref{eq:H} and Eq.~\eqref{eq:ffA1b}, while the phase $\eta=-1$ arise from the fermionic nature of the degrees of freedom. We refer to Ref.~\cite{schwarz2016lindblad,citro2018out} for a first-principle derivation of Eq.~\eqref{eq:modadj}. 

Quite interestingly we can write exact equations of motions for those Green's functions. These read
\begin{align}
	\frac{d}{dt} G^R(t) = -i \delta(t)\mathbf{1} - i \mathbb{H}_\mathrm{eff} G^R(t). \label{eq:eomR}
\end{align}
In frequency domain, the solution of Eq.~\eqref{eq:eomR} for the retarded function, and the advanced Green's function are given by
\begin{align}\label{eq:GR_GA}
	G^R(\omega)& = (\omega-\mathbb{H}_\mathrm{eff})^{-1},\\
	G^A(\omega)& = (G^R(\omega))^\dagger=  (\omega-\mathbb{H}^\dagger_\mathrm{eff})^{-1}.
\end{align}
The tridiagonal and homogeneous nature of the matrix $\mathbb{H}_\mathrm{eff}$ permits an exact solution for the component of $G^R(\omega)$~\cite{yueh2006explicit,yueh2008explicit}. Introducing the auxiliary variable
\begin{align}
	\phi = \arccos\left(-\frac{1}{2} \left(\omega + i \frac{\gamma}{2}\right)\right),\qquad \mathrm{Re}(\phi)\in [0,\pi],
\end{align}
the matrix element $G^R_{m,n}(\phi(\omega))$ with $m\ge n$ are
\begin{widetext}
\begin{align}
	G^R_{m,n}(\phi) & = \frac{\sin((L-m)\phi)}{\Delta(\phi)}\left(\frac{i \Gamma}{2} \sin(n\phi) + \frac{\Gamma^2}{4 J}\sin((n-1)\phi)\right) + \frac{\sin((L-m+1)\phi)}{\Delta(\phi)}\left(\frac{i \Gamma}{2} \sin((n-1)\phi) - J \sin(n\phi)\right),\label{eq:matel}\\
	\Delta(\phi) &= \sin(\phi) \left(J^2 \sin((L+1)\phi) - \frac{\Gamma^2}{4} \sin((L-1)\phi) - i J \Gamma \sin(L\phi)\right).\label{eq:delta}
\end{align}
\end{widetext}
The remaining elements are given by transposition, as the retarded Green function is symmetric $G^R_{m,n} = G^R_{n,m}$. A comparison between numerical exact diagonalization and the above analytic expressions is provided in Appendix~\ref{app:numcheck}, as well as the scaling limit expression of these functions.

From the retarded Green function, we find the solution of the Keldysh equation of motion in frequency space. We have
\begin{align}
	G^K(\omega) &= -2 i G^R(\omega) \Lambda G^A(\omega),\\
	\Lambda& \equiv P(\infty) - \mathbb{D}.\label{eq:keldyshfourier}
\end{align}
with $\Lambda$ the exact self-energy of the system, and $P(\infty)$ and $\mathbb{D}$ defined in Sec.~\ref{subsec:equal}. A derivation of this expression is given in Appendix~\ref{app:keldysh}.

\subsection{A relation between current and Green's Functions for the Driven-Dephasing Model}

Using the equation of motions technique we can obtain an exact relation between the stationary current and the Green's functions for the driven-dephasing model. This generalizes recent results for a model without dephasing~\cite{jin2020generic} analog to the Meir-Wingreen formula~\cite{meir1992landauer}, obtained within the Keldysh path integral formulations~\cite{kamenev2011nonequilibrium}.

We proceed as in Ref.~\cite{jin2020generic} and evaluate the stationary current at the boundary. As previously argued, since in our setting the current is constant throughout the system, it is convenient to use (either) expression in Eq.~\eqref{eq:edgecurr} to relate the current to the $R/A/K$ Green's functions by means of the density.

From the definition Eq.~\eqref{eq:nesskeldysh}, we have
	\begin{align}
		n_m(\infty) = \langle \hat{c}_n^\dagger & \hat{c}_m \rangle_\mathrm{NESS} = \nonumber \\
		& =\frac{i}{2}\int \frac{d\omega}{2\pi} (G^R_{m,m}(\omega) - G^A_{m,m}(\omega) - G^K_{m,m}(\omega)).\label{eq:nness}
	\end{align}
Using the sum rule, valid in the NESS,
\begin{align}
	\mathbf{1} = i\int \frac{d\omega}{2\pi} (G^R(\omega) - G^A(\omega)),
	\label{eq:retadvresol}
\end{align} 
and $j_\infty = \Gamma (1-n_1)$ we find after simple algebra (see also Appendix~\ref{app:keldysh} cfr. Eq~\eqref{eq:altkeld}) 
\begin{align}
	j_{\infty} = \Gamma \int \frac{d\omega}{2\pi} \langle 1| G^R(\omega) P(\infty) G^A(\omega)|1\rangle.\label{eq:finalres}
\end{align}
We conclude this section by remarking that, if the integral over frequency are formally performed, we recast the current read of Eq.~\eqref{eq:thetaf1}. In fact, from the eigendecomposition of $\mathbb{H}_\mathrm{eff}$, and using
\begin{align}
	\int \frac{d\omega}{2\pi} \frac{1}{(\omega - \lambda_p)(\omega+\lambda_q)} = -\frac{1}{i\lambda_p -i \lambda^*_q}.
\end{align}
we recast 
\begin{align}
	j_\infty = \Gamma \sum_k \Theta_{1,1,k} p_k,
\end{align}
which is obtained from Eq.~\eqref{eq:finalres} and Eq.~\eqref{eq:edgecurr}. The advantage of Eq.~\eqref{eq:finalres} over Eq.~\eqref{eq:thetaf1} lies in the direct physical interpretation of the Green's functions, which are furthermore relevant also for experimental setups.

\subsection{Connection with Keldysh Field Theory and Stochastic Unraveling}

The results reported above show that using Lindblad equations of motion one can compute exactly the single particle Green's functions of the driven-dephasing model, despite of the fact that the Lindbladian is not quadratic in the fermionic creation/annihilation operators. This is somewhat remarkable from the point of view of the Keldysh action associated to the Lindblad master equation, since in this Keldysh framework the dephasing $\gamma$ would give rise to a dissipative \emph{interaction} vertex. Our results show that the associated diagrammatics should contains a number of non-trivial cancellations, such that a closed form self-energy could be written for the Green's functions. The resummabiity of the dephasing diagrammatics in the Keldysh language can be perhaps more easily understood by unravelling the Lindblad master equation into a stochastic quantum trajectory. In this picture in fact the noisy Hamiltonian would only features quadratic terms in the fermion operators and one could expect to be able to compute exactly the Green's function for this problem, at least for a given realization of the stochastic process. Averaging over the disorder would lead again to a diagrammatic expansion, which in light of our exact results within the Lindblad master equation will likely take a particularly simple form.  We have not attempted here to make progress in these direction for the driven-dephasing model, although recent results on related models~\cite{dolgirev2020nongaussian,jin2021inprep} seem to indicate this to be the case. In particular one can show that a Lindblad master equation is equivalent to a diagrammatic expansion in the system-bath (or noise) coupling and an exact resummation of non-crossing diagrams~\cite{scarlatella2021inprep}.

\section{Results}
\label{sec:results}
In this section we discuss our results for the current and the Green's functions of the model. For simplicity we fix $J=1$ in the following and study the properties of the system as a function of the dephasing $\gamma$ and the strength of the drive $\Gamma$. 

\subsection{Transport: current and density profile}
\label{subsec:transport}
\begin{figure}[t]
	\includegraphics[width=\columnwidth]{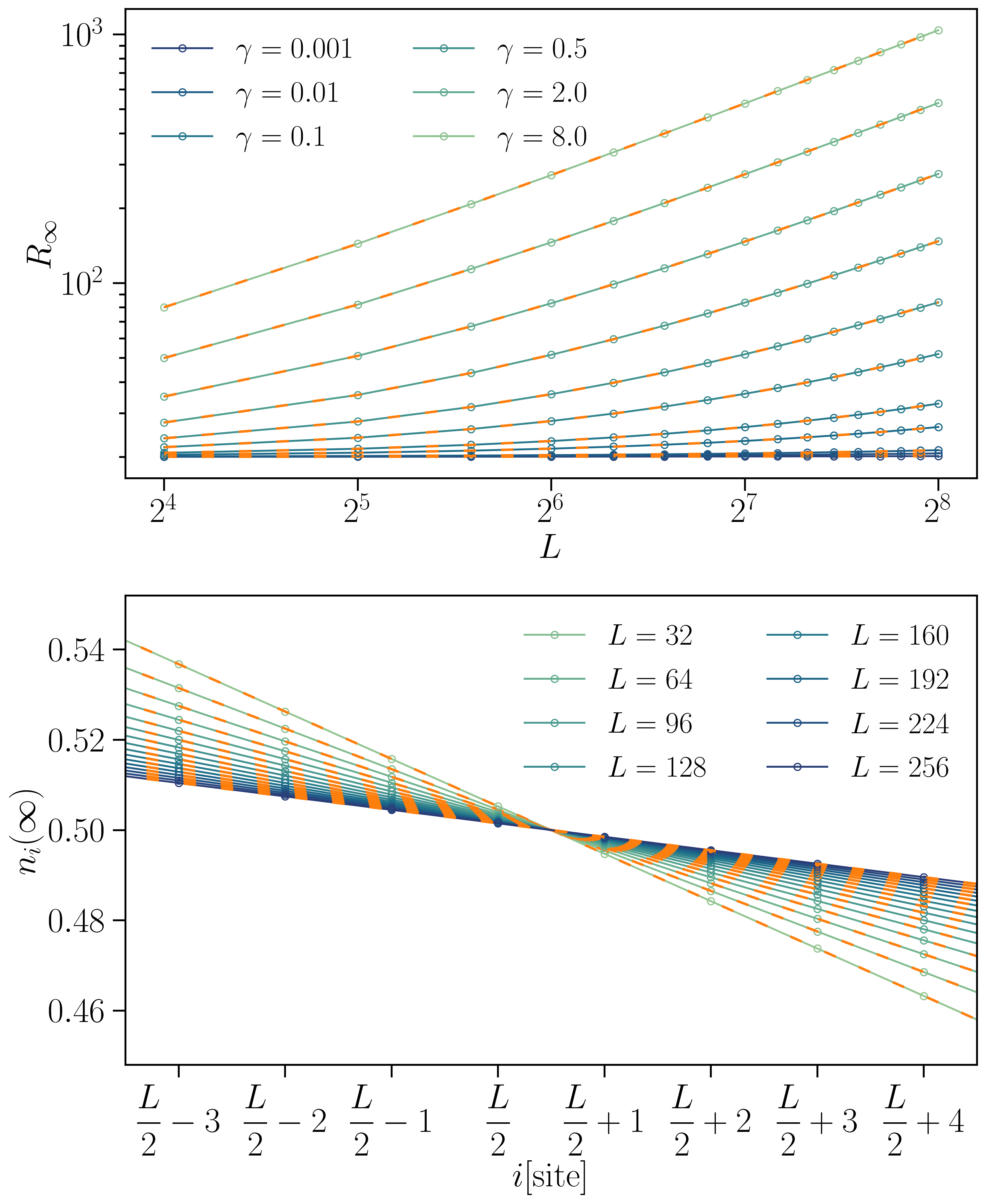}
	\caption{ (Top) Resistance of the system versus system size $L$ for different values of $\gamma$ and $\Gamma=0.1$. At small values of $\gamma$, the system clearly display a crossover between a ballistic and a diffusive regime. The dashed orange lines correspond to Eq.~\eqref{eq:znida2}.
	(Bottom) Density profile $n_i(\infty)$ for various system sizes $L$ for $\gamma=0.5$ and $\Gamma=0.1$. The slope around $i=L/2$ is $\propto 1/L$. The dashed orange lines correspond to Eq.~\eqref{eq:znida}. }\label{fig1:current_density}
\end{figure}

We start reviewing the transport properties of the system, which have been extensively discussed in the literature~\cite{znidaric2010exact,znidaric2010a,bertini2020finitetemperature}. In presence of a finite pump-loss boundary drive, a homogeneous stationary state with a finite current arises. We define the resistance $R_\infty = 1/j_{\infty}$.
In our setting, the stationary current is given by~\cite{znidaric2010exact}
\begin{align}
	j_\infty = \frac{4 \Gamma }{\Gamma^2 + 4 + (L-1)\Gamma\gamma}\simeq D\frac{4}{L}, \qquad D=1/\gamma\nonumber \\
	R_\infty = \frac{1}{2}\left(\frac{\Gamma}{2} + \frac{2}{\Gamma} + (L-1)\frac{\gamma}{2}\right) \simeq \frac{\gamma}{4} L.\label{eq:znida2}
\end{align}
Diffusive behavior is signaled by $R(L)\propto L$, as clearly shown in Fig.~\ref{fig1:current_density} (top panel). 
The existence of diffusion in the model without boundary drive can be established analytically by analyzing the spectral properties of the Lindbladian ~\cite{esposito2005emergence,esposito2005exactly}. We note that the resistance shows a crossover from a ballistic to a diffusive scaling  ~\cite{eisler2011crossover}, as a function of the dephasing rate $\gamma$ as we show in Fig.~\ref{fig1:current_density}.
A diffusive current should correspond to the onset of a stationary state density profile at site $i$ of the form~\cite{znidaric2010exact}
\begin{equation}
	\frac{\Delta n_i}{\Delta i} = \frac{\gamma \Gamma}{\Gamma^2+ 1+ (L-1)\gamma \Gamma}\rightarrow_{L\gg 1} \frac{1}{L}.\label{eq:znida}
\end{equation}

In the previous equation we have considered the scaling limit, where Fick's law is recovered. 
We see that the density drop near the boundary, while in the bulk the system in the thermodynamic limit approaches a constant density profile with $n\rightarrow 1/2$, consistent with an infinite temperature thermalization as we will discuss in more detail below.
\begin{figure}[t]
	\includegraphics[width=\columnwidth]{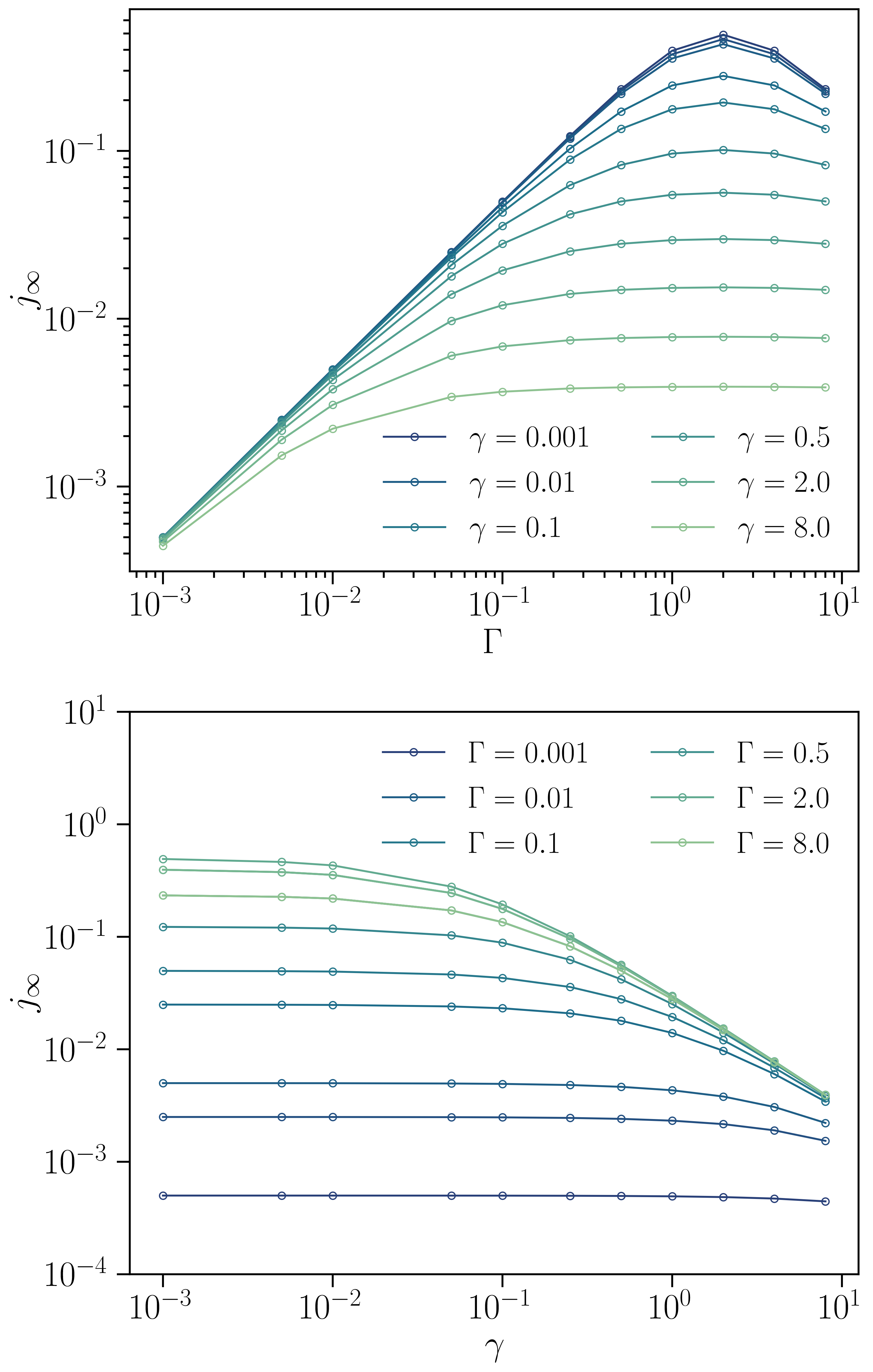}
	\caption{\label{fig2:nonlinear_transport} Stationary current $j_\infty$ as a function of the drive strength $\Gamma$ (top panel) and dephasing rate $\gamma$ (bottom panel) for a fixed system size $L=256$. We see that the linear regime $j_\infty\sim \Gamma$ is independent of the dephasing rate while the maximum current is strongly suppressed by increasing $\gamma$, which plays a role similar to inelastic many-body interactions.}
\end{figure}

We now discuss the dependence of the stationary current $j_\infty$ from the strength of the drive $\Gamma$ and the dephasing rate $\gamma$. As we show in the top panel of Fig.~\ref{fig2:nonlinear_transport}, the current is linear at small $\Gamma$, reaches a maximum at $\Gamma^*$ and then starts to decrease, eventually going to zero as $1/\Gamma$ for $\Gamma\gg1$. This indicates that when the drive is much larger than the single particle bandwidth resonance responsible for coherent transport is not efficiently established and the current decreases. A finite dephasing rate does not change the linear regime while strongly reduces the maximum current which can be sustained by the system. From Eq.~(\ref{eq:znida2}) we obtain that $j^*_\infty$, corresponding to the current at the optimal coupling $\Gamma^*$, scales as $j^*_\infty\sim 1/\gamma L $. In the bottom panel we show the current as a function of the dephasing rate from which again we can see how the linear regime is almost unaffected by the scattering while at stronger drive the current decreases with $\gamma$ and eventually vanishes as $1/\gamma$ in accordance with Eq.~(\ref{eq:znida2}). Overall these results confirm the physical picture that a finite dephasing rate is detrimental to coherent transport, playing a similar role as inelastic scattering due to many-body interactions or heating due to finite temperature.

\subsection{Local density of states and distribution function}
We now discuss the Green's functions of the model, which to the best of our knowledge have not been presented before. We start with the retarded Green's function at site $n$, whose imaginary part
\begin{equation}
	A_n(\omega)=-\frac{1}{\pi}\mbox{Im}G^R_{n,n}(\omega)
\end{equation}
describes the local Density of State (DoS) of the system. This can be obtained directly from the results of Sec.~\ref{subsec:green} (cfr. Eq.~\eqref{eq:matel}). 

Before discussing the case $\gamma>0$, we briefly comment on the $\gamma=0$ case. In this situation, Eq.~\eqref{eq:matel} reduces to ballistic formula, and the results in Ref.~\cite{jin2020generic} are recovered. 

Turning on the dephasing in the system has important consequences. The auxiliary variable $\phi = \arccos\left(-\frac{1}{2} \left(\omega + i \frac{\gamma}{2}\right)\right)$ acquire an imaginary part which suppresses the matrix elements $G_{m,n}^R$. Quantitatively, starting from Eq.~\eqref{eq:matel} we obtain the scaling limit $L\gg1$
\begin{align}
	G^R_{m,n}(\phi) = -e^{i m\phi} \frac{2 i \sin(n\phi) + \Gamma \sin((n-1)\phi)}{\sin(\phi) (2 i + e^{i\phi}\Gamma)}.\label{eq:scalingr}
\end{align}
The above expression is exact for the thermodynamic limit of any component of the retarded function, provided $\gamma>0$. 
For a site in the bulk $n\sim L/2\gg 1$ the above expression simplifies to
\begin{align}
	G^R_{n,n}(\phi) = \frac{-i}{\sqrt{4-\left(\omega+i\gamma/2\right)^2}}\label{eq:thermoscale}
\end{align}
which as expected is independent both from the site index and from the boundary drive $\Gamma$. Importantly, Eq.~\eqref{eq:thermoscale} matches the translational invariant result (Appendix~\ref{app:trans}), see Appendix~\ref{app:numcheck} for a comparison.

In Fig.~\ref{fig3:localdos} we plot the local DoS in the middle of the chain and at the boundary (respectively top and bottom panels) for different values of $\gamma$. In the bulk we see for small dephasing two sharp edges around $\omega=\pm 2$ which are further smeared out by increasing the dephasing rate. We can understand this result by noticing that in absence of any boundary drive the system is translational invariant (see Appendix~\ref{app:trans}) and the local DoS displays the characteristic square-root singularities at the edge of the single particle bandwidth smeared by a Lorentzian weight which is due to the dephasing and that results into
additional tails at large frequencies. As expected therefore the behavior in the bulk is essentially independent of the value of the boundary drive, at least for large enough system sizes. Close to the boundary on the other hand we expect a much stronger dependence from the drive strength, as we show in Appendix~\ref{app:numcheck}.
  In the bottom panel of Fig.~\ref{fig3:localdos}  we plot the boundary local DoS for different values of the dephasing rate, showing that while the single particle band edges are still visible for small $\gamma$ there are no signatures of the sharp singularities seen in the bulk. For large values of $\gamma$ instead the bulk and the boundary DoS are essentially the same, corresponding to very broad line width.
  
  We see that despite the \textit{interacting} nature of the dephasing rate, which enters the Lindblad master equation with a quadratic jump operator, the scale $\gamma$ play essentially the role of a dissipative width and does not induce any frequency dependent self-energy corrections for the momentum resolved Green's function, while the local DoS has a sizeable spectral redistribution from low to high frequency due to the dephasing.

\begin{figure}[t]
	\includegraphics[width=\columnwidth]{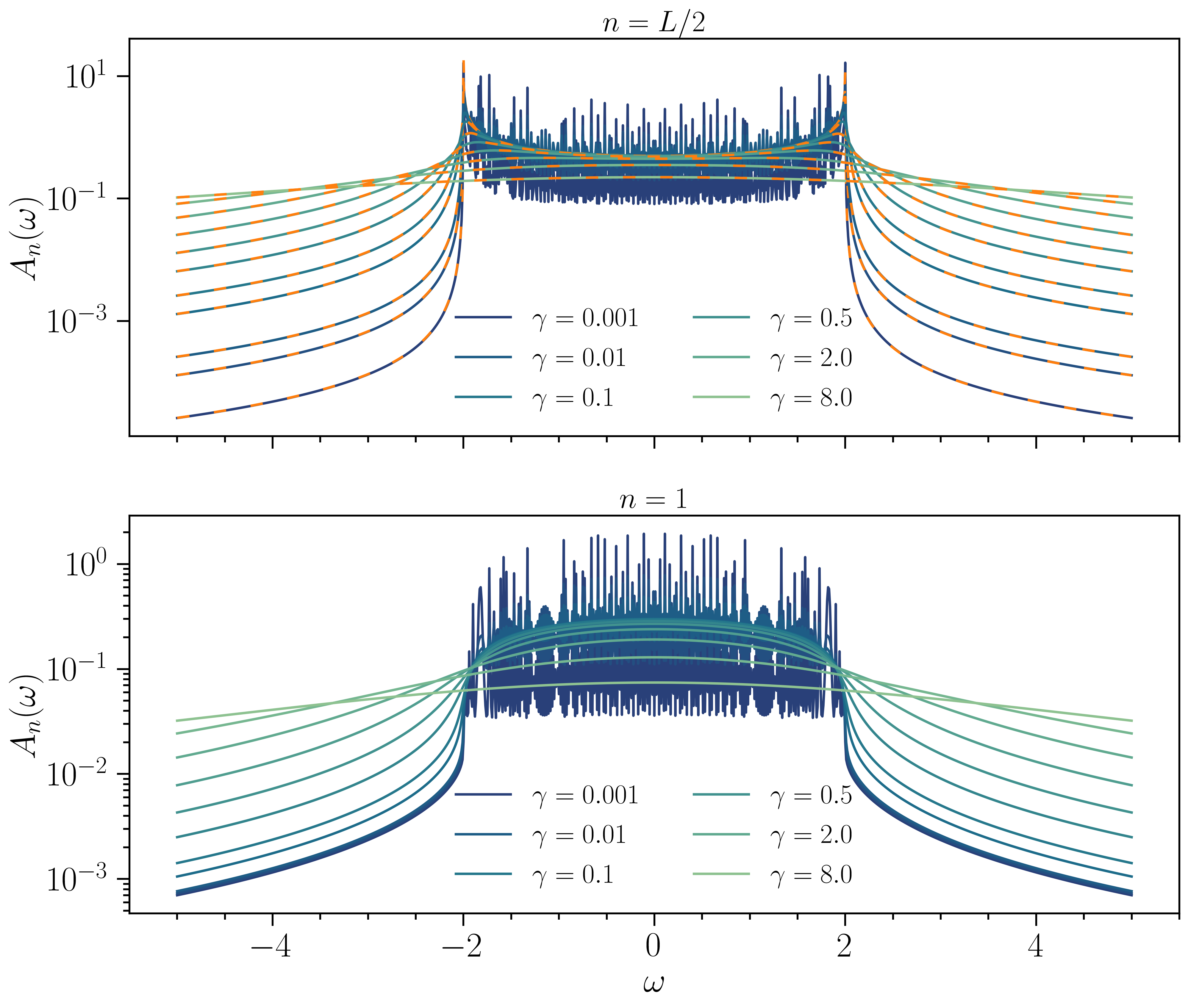}
	\caption{\label{fig3:localdos} Local Density of States. (Top) $A_{n}(\omega)$ at the center of the chain $n=L/2$ as a function of frequency, compared with the translational invariant result Eq.~\eqref{eq:thermoscale} (dashed orange lines), for different values of the dephasing. (Bottom) Local DoS $A_{n}(\omega)$ at the boundary of the chain $n=1$  as a function of frequency and different values of the dephasing rate.}
\end{figure}

A complementary perspective on the competition between dephasing and boundary terms is provided by the local Keldysh Green's function $G_n^{K}(\omega)$, defined in Eq.~\eqref{eq:defkeld}, which describe the occupation of the single particle excitations on top of the stationary state. It is convenient to parametrize it through the distribution function, which is defined in analogy with the thermal equilibrium case.

\begin{figure}[t]
	\includegraphics[width=\columnwidth]{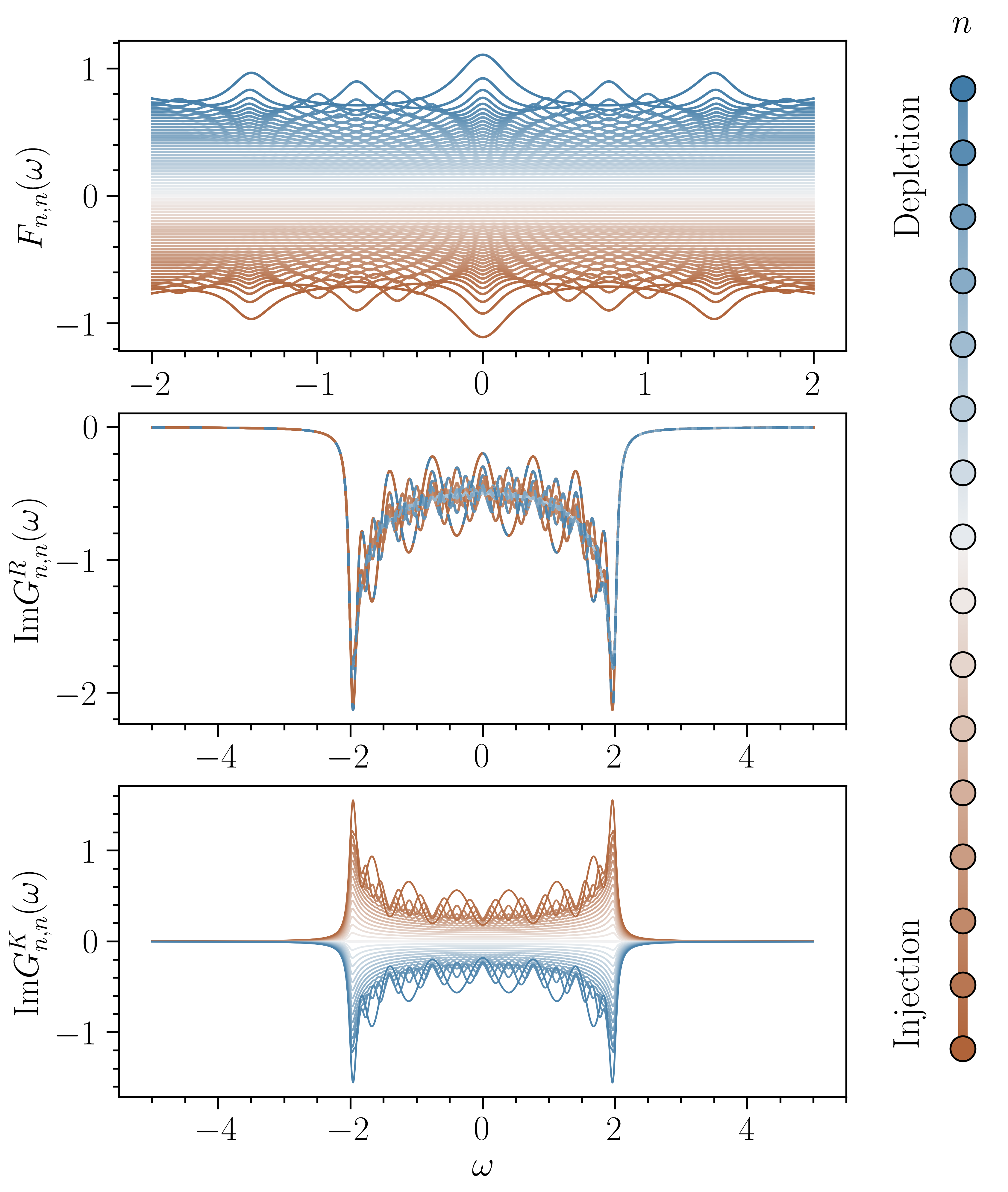}
	\caption{ \label{fig4:F_neq} For all panels we fix $L=256$, $\gamma=0.1$, and $\Gamma=0.1$. (Top) Distribution function at different sites in the chain. (Center) Imaginary part of the retarded function. (Bottom) Imaginary part of the Keldysh Green's function.
	}
\end{figure}
In fact, if the system was in true thermal equilibrium, the quantum fluctuation-dissipation theorem (FDT) would constrain the Keldysh and the retarded components to obey the relation~\cite{kamenev2011nonequilibrium}
\begin{equation}
\label{eq:fdt}
\frac{G_{n,n}^{K}(\omega)}{-2 \pi i A_{n,n}(\omega)} \equiv F^{\rm eq}(\omega)= \tanh ( \frac{ \omega }{2 T } ) 
\end{equation}
where $T$ is the system temperature. At low frequency or high temperatures $\omega \ll T$, one has $F^{\rm eq}(\omega)\sim \omega/T$. In a non-equilibrium system, on the contrary, there is no well defined temperature and the FDT does not hold in general. Nonetheless, it is useful to use the left-hand side of the FDT in Eq.~\eqref{eq:fdt} to \textit{define} an effective distribution function
\begin{align}
	F^{\rm neq}_n(\omega)=\frac{iG_{n,n}^{K}(\omega)}{2 \pi A_{n,n}(\omega) }\label{eq:distrib}.
\end{align}
We plot the non-equilibrium distribution $F^{\rm neq}_n(\omega)$ in Fig.~\ref{fig4:F_neq} for different lattice sites, (corresponding to different lines in each panel) and we compare it to the retarded and Keldysh Green's functions (see central and bottom panels). We see that the distribution function close to the boundaries (top and bottom lines)  is strongly frequency dependent and highly athermal. In particular $F^{\rm neq}_n(\omega)$  is an even function of frequency, unlike the thermal result in Eq.~(\ref{eq:fdt}), and odd under spatial inversion around the center of the lattice. This feature arises directly from the Keldysh Green's function (see bottom panel of Fig.~\ref{fig4:F_neq}) while the retarded Green's function is symmetric under inversion (central panel), and reflects the symmetry between injection and depletion at the boundaries. In fact we can see from Eq.~(\ref{eq:nness}) that the frequency integral over the Keldysh Green's function quantifies the deviation of the local occupation from the bulk value $n=1/2$. Therefore near the boundary where injection(depletion) occurs we expect an increase(decrease) of density and correspondingly a Keldysh Green's function which is of positive(negative) sign. Alternatively, we can parametrize the distribution function in term of an effective Fermi function $F^{\rm neq}_n(\omega)=1-2f_n(\omega)$ from which conclude that positive(negative) values of the distribution function correspond to an increased(decreased) occupancy of the Fermi function.

\begin{figure}[t!]
	\includegraphics[width=\columnwidth]{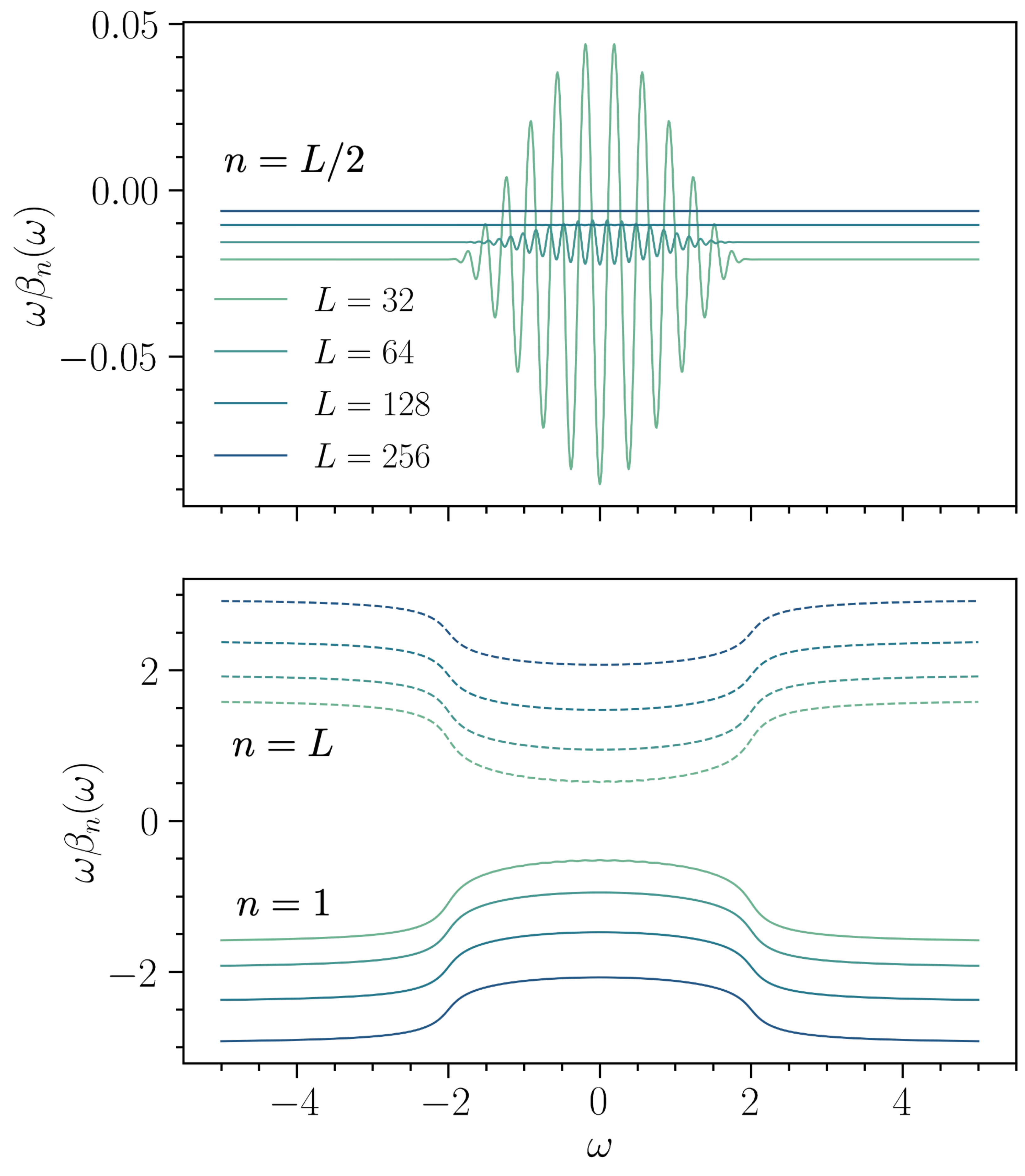}
	\caption{\label{fig6} Inverse effective temperature $\omega \beta_n(\omega)$, defined as in Eq.(\ref{eq:betaeff}) of the main text, at $\Gamma=1/4$ and $\gamma=1/4$ (Top) in the bulk and (Bottom) in the boundaries, for different system sizes. }
\end{figure}
Upon approaching the bulk of the system, on the other hand, we see that the distribution function becomes flatter in frequency and its values approaches zero, a signature of the onset of bulk thermalization to infinite temperature. To further discuss this point we introduce a frequency-dependent inverse effective temperature~\cite{MitraEtAlPRL06,ClerkRMP2010,FoiniCugliandoloGambassi_PRB11,DallaTorreetalPRB2012,SchiroMitraPRL14,SiebererRepProgPhys2016} 
\begin{equation}\label{eq:betaeff}
	\beta_n (\omega) = \frac{1}{2\omega}\mathrm{arctanh}(F^\mathrm{ neq}_n(\omega))
\end{equation} 
We plot the inverse effective temperature in Fig.~\ref{fig6} at the bulk and at the boundaries for different system sizes $L$.  We see that in the bulk of the system ($n=L/2$, top panel) $\beta_n (\omega)$ shows small oscillations for frequencies within the single particle bandwidth, which are rapidly damped out upon increasing $L$ leaving a constant value which approaches zero in the thermodynamic limit, with a uniform scaling $1/L$ that comes from the Keldysh Green's function. At the boundary instead the effective temperature depends on frequency, as we show in the bottom panel of Fig.~\ref{fig6}. In particular we note that for frequencies within the single particle bandwidth $\beta_n (\omega)$ is smaller in absolute value, corresponding to an increased heating and thermalization due to resonant single particle processes as compared to the high frequency regions.

\section{Conclusion}
\label{sec:conclusions}

In this paper we have studied transport and thermalization in a simple model of non-interacting one dimensional spinless fermions subject to local dephasing and boundary drive. Within the framework of the Lindblad master equation of open quantum systems we have shown that the problem can be exactly solved both for what concerns single particle static correlation functions, giving access to the current in the non-equilibrium steady state,  as well as for the single particle Green's functions describing excitations on top of the stationary state.
In particular we have obtained a closed form expression for the local retarded Green's function, from which the Keldysh component can be readily obtained, and a relation between the stationary current and the single particle Green's functions of the system, similar to the famous Meir-Wingreen formula, which generalizes to the dephasing case a recent result obtained within Keldysh techniques.

Using this framework we have revisited the transport properties of the system, featuring a well studied crossover between ballistic and diffusive behavior in the scaling of the current versus system size. We have discussed how a finite dephasing rate introduces a resistivity growing with system size and a density profile decreasing linearly in the bulk. Going beyond the linear transport regime we have further shown that the dephasing strongly reduces the maximum current which can flow through the system at a given system size. 

Furthermore we discussed the spectral properties of the model as encoded in the local Density of States and in the single particle distribution function. We have shown that the former in the bulk  can be understood by considering the homogeneous translational invariant case, where dephasing plays essentialy the role of an additional linewidth broadening the edge singularities of the single particle DoS, while the boundary DoS features a much broader linewidth due to the additional driving term. Finally we have shown that the distribution function at the boundary is highly non-equilibrium, reflecting the local injection and depletion of particles, while approaching the bulk site it becomes flatter in frequency consistent with the onset of infinite temperature thermalization.

We conclude by noticing that the above analysis has been performed for a translational invariant system (boundary terms aside), however the Lindblad equation of motion technique used here can be naturally extended to disordered systems where one expect anomalous transport behavior. Results along these lines will be reported elsewhere~\cite{turkeshi2021inprep}.

\begin{acknowledgements}
We thank D. Barbier, L. Cugliandolo, and M. Tarzia for collaboration and discussions on related topics.
We acknowledge computational resources on the Coll\'ege de France IPH cluster. The work of XT is partly supported by the ERC under grant number 758329 (AGEnTh), by the MIUR Programme FARE (MEPH), and has received funding from the European Union’s Horizon 2020 research and innovation programme under grant agreement No 817482.
This work was supported by the ANR grant ''NonEQuMat'' (ANR-19-CE47-0001).
\end{acknowledgements}

\appendix
\section{Spectrum of the non-Hermitian Hamiltonian}
\label{app:heff}
In this section we briefly review the diagonalization of the non-Hermitian Hamiltonian $\mathbb{H}_\mathrm{eff}$ by means of coordinate Bethe Ansatz. A throughout analysis on the existence of the solutions and on their derivation given in Ref.~\cite{yueh2006explicit,yueh2008explicit}. For simplicity we discuss the case of the right eigenvector, and fix $J=1$.
Using the Bethe Ansatz $\varphi_{l}(k) = A_k e^{i \theta_k l} + B_k e^{-i\theta_k l}$,  the eigensystem 
\begin{align}
	\mathbb{H}_\mathrm{eff} |\varphi(k)\rangle = \lambda_k |\varphi(k)\rangle
\end{align}
reduces to 
\begin{align}
	A_k &= - \frac{2 i+\Gamma e^{-i\theta_k}}{2 i+\Gamma e^{i\theta_k}} B_k\label{eq:ratioab}\\
	\lambda_k &= -i \frac{\gamma}{2} + 2 \cos \theta_k,\\
	0 &= \sin((L+1)\theta_k ) - \frac{\Gamma^2}{4} \sin((L-1)\theta_k) -i \Gamma \sin (L\theta).\label{eq:quantization}
\end{align}
The quantization condition Eq.~\eqref{eq:quantization} fixes the values of $\theta_k\in[0,\pi[$, and hence the eigenvalues $\lambda_k$. However, for generic values of $\Gamma$, we lack a closed form expression of the complex angles $\theta_k$, which need to be computed analytically. Inserting the obtained expressions for $\theta$ in the definition of $|\varphi(k)\rangle$ and using Eq.~\eqref{eq:ratioab}, we obtain the right eigenvector. Similarly the left eigenvector the be obtained through symmetry argument on $\mathbb{H}_\mathrm{eff}$. The norm (free coefficient $B_k$) is fixed by the biorthonormal condition.

\section{Derivation of the Keldysh Green's function}
\label{app:keldysh}
The equation of motion of the Keldysh Green's function is 
\begin{align}
	G^K(t) = \begin{cases}
		e^{tT} G^K(0), & t>0,\\
		G^K(0)e^{-tT^\dagger} , & t<0.
	\end{cases}
	\label{eq:solutionkeldysh}
\end{align}
In frequency domain we have
\begin{align}
	G^K(\omega) &= \int_{-\infty}^{\infty} dt e^{i\omega t} G^K(t)\nonumber\\
	& = \int_0^\infty dt e^{i\omega t + tT} G^K(0) + \int_{-\infty}^0 dt G^K(0) e^{i\omega t -t T^\dagger}\nonumber\\
	& = \frac{-1}{i(\omega-iT)} G^K(0) + G^K(0)\frac{1}{i(\omega + i T^\dagger)}\nonumber \\
	& = \frac{i}{\omega-iT} G^K(0) - G^K(0)\frac{i}{\omega+iT^\dagger}.
\end{align} 
From the definition Eq.~\eqref{eq:defkeld}, we read $G^K(0) = i - 2i C(\infty)$. Hence
\begin{align}
	G^K(\omega) &= - G^R(\omega) + G^A(\omega) +\nonumber\\&+ 2G^R(\omega)\left(iC(\infty) T^\dagger + iT C(\infty)\right) G^A(\omega)\\
	& = - G^R(\omega) + G^A(\omega) +\nonumber\\
	& \qquad - 2 i G^R(\omega) P(\infty) G^A(\omega).\label{eq:altkeld}
\end{align}
The final expression Eq.~\eqref{eq:keldyshfourier} is recovered once the difference between retarded and advanced Green functions is performed
\begin{align}
	- G^R(\omega) + G^A(\omega) = 2 i G^R(\omega) \mathbb{D} G^A(\omega).
\end{align}
\textit{En passant}, we note expression Eq.~\eqref{eq:altkeld} is useful in deriving the Landauer formula Eq.~\eqref{eq:finalres}.

\section{Translational invariance pure dephasing model}
\label{app:trans}
If we disregard the boundary drive, and close the chain to enforce periodic boundary conditions, the system becomes translational invariant. 
In this situation, the correlation functions depend only of the distance, and the equations of motion for the Green's functions can be solved in Fourier space. 
We define the momentum resolved Green's functions
\begin{align}
G^{R/A/K}(q,\omega)=\sum_n e^{-i nq}G^{R/A/K}_{n}(\omega)
\end{align}
where we have introduced the translational invariant Green's functions $G^{R/A/K}_{m,n}(\omega)=G^{R/A/K}_{m-n}(\omega)$. The retarded component takes a simple form 
\begin{align}
G^{R}(q,\omega)=\frac{1}{\omega-\varepsilon(q)-i\gamma/2}
\end{align}
which gives a spectral function in the form of a lorentzian centered around the bare dispersion $\varepsilon(q)=-2J\cos(2 \pi q/N)$. The local density of states become site independent and equal to
\begin{align}
A(\omega)&=-\frac{1}{\pi}\sum_q \frac{2 \gamma}{4\left(\omega-\varepsilon(q)\right)^2+\gamma^2}.
\end{align}
We introduce the density of states of the one dimensional tight binding chain
\begin{align}
	D(\varepsilon)&=\sum_q\delta(\varepsilon-\varepsilon(q)) = \int_{-\pi}^\pi dq \delta(\varepsilon+2J \cos(q)) \nonumber \\
	&= \theta(4 J -\varepsilon^2)
 	\frac{1}{\sqrt{4J -\varepsilon^2}},
\end{align} 
which is centered around $\varepsilon=0$ with a bandwidth of $4J$ and sharp edges. The smear by dephasing results into states in the tails outside of the kinetic bandwidth.
With the above definition the final result for the local density of state is 
\begin{align}
	A(\omega)& =-\frac{1}{\pi}
\int d\varepsilon D(\varepsilon)\frac{2 \gamma}{ 4\left(\omega-\varepsilon\right)^2+\gamma^2} \nonumber\\ &=-\frac{2\gamma}{\pi}
\int_{-2 J}^{2J} d\varepsilon \frac{1}{\sqrt{4 J-\varepsilon^2}}\frac{1}{4\left(\omega-\varepsilon\right)^2+\gamma^2}.\label{eq:aaa}
\end{align}

\begin{figure}[t!]
	\includegraphics[width=\columnwidth]{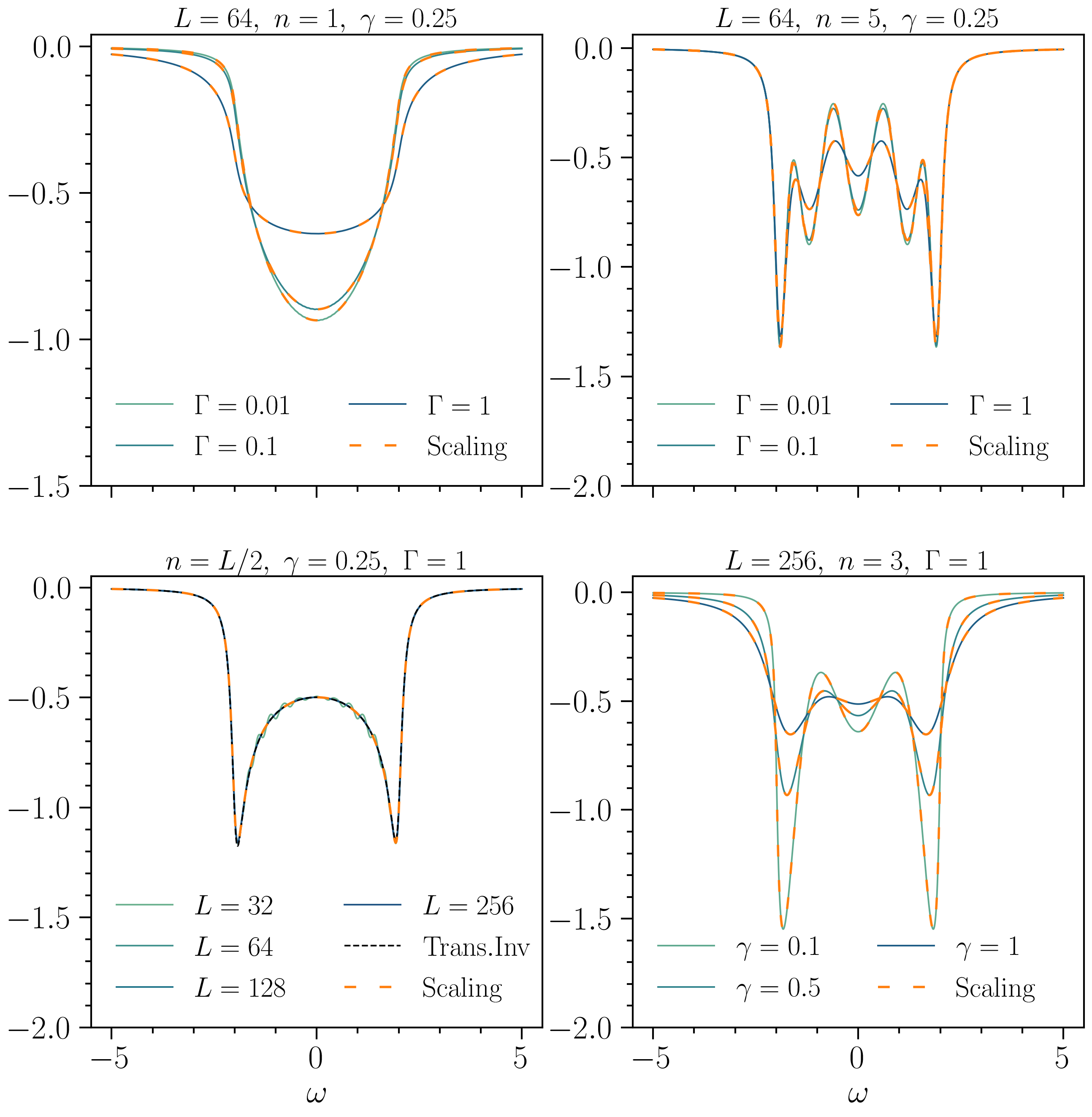}
	\caption{\label{fig:comparison} Comparison between Eq.~\eqref{eq:matel} and the exact diagonalization results from Eq.~\eqref{eq:retadvresol}. The dashed orange lines are Eq.~\eqref{eq:matel}, whereas the dotted black line is the translational invariant limit Eq.~\eqref{eq:aaa}. }
\end{figure}

The Keldysh Green's function on the other hand can be obtained from the Eq.~\eqref{eq:keldyshfourier}. Because of translational symmetry, $\Lambda = \gamma (1-2 n)/2$ is proportional to the identity. Hence, the Keldysh Green's function in Fourier space reads
\begin{equation}
	G^K(q,\omega)=-i \gamma \left(1-2n\right) G^R(q,\omega)G^A(q,\omega),
\end{equation}
and it is therefore identically zero, since $n=1/2$, which is again consistent with infinite temperature thermalization.

\section{Numerical benchmarks}
\label{app:numcheck}
We benchmark the exact retarded Green function Eq.~\eqref{eq:matel} with the results from exact diagonalization and the values of the translational invariant system. 
As in Sec.~\ref{sec:results}, we fix $J=1$. The comparison is given in Fig.~\ref{fig:comparison}.
We appreciate the formula matches exactly the results for different sites, different parameters $\gamma$ and $\Gamma$ and matches the translational invariant limit in the bulk. 
\bibliography{diffusion}

\begin{thebibliography}{39}%
\makeatletter
\providecommand \@ifxundefined [1]{%
 \@ifx{#1\undefined}
}%
\providecommand \@ifnum [1]{%
 \ifnum #1\expandafter \@firstoftwo
 \else \expandafter \@secondoftwo
 \fi
}%
\providecommand \@ifx [1]{%
 \ifx #1\expandafter \@firstoftwo
 \else \expandafter \@secondoftwo
 \fi
}%
\providecommand \natexlab [1]{#1}%
\providecommand \enquote  [1]{``#1''}%
\providecommand \bibnamefont  [1]{#1}%
\providecommand \bibfnamefont [1]{#1}%
\providecommand \citenamefont [1]{#1}%
\providecommand \href@noop [0]{\@secondoftwo}%
\providecommand \href [0]{\begingroup \@sanitize@url \@href}%
\providecommand \@href[1]{\@@startlink{#1}\@@href}%
\providecommand \@@href[1]{\endgroup#1\@@endlink}%
\providecommand \@sanitize@url [0]{\catcode `\\12\catcode `\$12\catcode
  `\&12\catcode `\#12\catcode `\^12\catcode `\_12\catcode `\%12\relax}%
\providecommand \@@startlink[1]{}%
\providecommand \@@endlink[0]{}%
\providecommand \url  [0]{\begingroup\@sanitize@url \@url }%
\providecommand \@url [1]{\endgroup\@href {#1}{\urlprefix }}%
\providecommand \urlprefix  [0]{URL }%
\providecommand \Eprint [0]{\href }%
\providecommand \doibase [0]{https://doi.org/}%
\providecommand \selectlanguage [0]{\@gobble}%
\providecommand \bibinfo  [0]{\@secondoftwo}%
\providecommand \bibfield  [0]{\@secondoftwo}%
\providecommand \translation [1]{[#1]}%
\providecommand \BibitemOpen [0]{}%
\providecommand \bibitemStop [0]{}%
\providecommand \bibitemNoStop [0]{.\EOS\space}%
\providecommand \EOS [0]{\spacefactor3000\relax}%
\providecommand \BibitemShut  [1]{\csname bibitem#1\endcsname}%
\let\auto@bib@innerbib\@empty
\bibitem [{\citenamefont {Bertini}\ \emph {et~al.}(2015)\citenamefont
  {Bertini}, \citenamefont {De~Sole}, \citenamefont {Gabrielli}, \citenamefont
  {Jona-Lasinio},\ and\ \citenamefont {Landim}}]{bertini2015macroscopic}%
  \BibitemOpen
  \bibfield  {author} {\bibinfo {author} {\bibfnamefont {L.}~\bibnamefont
  {Bertini}}, \bibinfo {author} {\bibfnamefont {A.}~\bibnamefont {De~Sole}},
  \bibinfo {author} {\bibfnamefont {D.}~\bibnamefont {Gabrielli}}, \bibinfo
  {author} {\bibfnamefont {G.}~\bibnamefont {Jona-Lasinio}},\ and\ \bibinfo
  {author} {\bibfnamefont {C.}~\bibnamefont {Landim}},\ }\bibfield  {title}
  {\bibinfo {title} {Macroscopic fluctuation theory},\ }\href
  {https://doi.org/10.1103/RevModPhys.87.593} {\bibfield  {journal} {\bibinfo
  {journal} {Rev. Mod. Phys.}\ }\textbf {\bibinfo {volume} {87}},\ \bibinfo
  {pages} {593} (\bibinfo {year} {2015})}\BibitemShut {NoStop}%
\bibitem [{\citenamefont {Bertini}\ \emph {et~al.}(2021)\citenamefont
  {Bertini}, \citenamefont {Heidrich-Meisner}, \citenamefont {Karrasch},
  \citenamefont {Prosen}, \citenamefont {Steinigeweg},\ and\ \citenamefont
  {\ifmmode \check{Z}\else \v{Z}\fi{}nidari\ifmmode~\check{c}\else
  \v{c}\fi{}}}]{bertini2020finitetemperature}%
  \BibitemOpen
  \bibfield  {author} {\bibinfo {author} {\bibfnamefont {B.}~\bibnamefont
  {Bertini}}, \bibinfo {author} {\bibfnamefont {F.}~\bibnamefont
  {Heidrich-Meisner}}, \bibinfo {author} {\bibfnamefont {C.}~\bibnamefont
  {Karrasch}}, \bibinfo {author} {\bibfnamefont {T.}~\bibnamefont {Prosen}},
  \bibinfo {author} {\bibfnamefont {R.}~\bibnamefont {Steinigeweg}},\ and\
  \bibinfo {author} {\bibfnamefont {M.}~\bibnamefont {\ifmmode \check{Z}\else
  \v{Z}\fi{}nidari\ifmmode~\check{c}\else \v{c}\fi{}}},\ }\bibfield  {title}
  {\bibinfo {title} {Finite-temperature transport in one-dimensional quantum
  lattice models},\ }\href {https://doi.org/10.1103/RevModPhys.93.025003}
  {\bibfield  {journal} {\bibinfo  {journal} {Rev. Mod. Phys.}\ }\textbf
  {\bibinfo {volume} {93}},\ \bibinfo {pages} {025003} (\bibinfo {year}
  {2021})}\BibitemShut {NoStop}%
\bibitem [{\citenamefont {Breuer}\ and\ \citenamefont
  {Petruccione}(2007)}]{breuerPetruccione2010}%
  \BibitemOpen
  \bibfield  {author} {\bibinfo {author} {\bibfnamefont {H.~P.}\ \bibnamefont
  {Breuer}}\ and\ \bibinfo {author} {\bibfnamefont {F.}~\bibnamefont
  {Petruccione}},\ }\href
  {https://doi.org/10.1093/acprof:oso/9780199213900.001.0001} {\emph {\bibinfo
  {title} {The {{Theory}} of {{Open Quantum Systems}}}}},\ \bibinfo {edition}
  {1st}\ ed.,\ Vol.\ \bibinfo {volume} {9780199213}\ (\bibinfo  {publisher}
  {{OUP Oxford}},\ \bibinfo {year} {2007})\BibitemShut {NoStop}%
\bibitem [{\citenamefont {Esposito}\ and\ \citenamefont
  {Gaspard}(2005{\natexlab{a}})}]{esposito2005exactly}%
  \BibitemOpen
  \bibfield  {author} {\bibinfo {author} {\bibfnamefont {M.}~\bibnamefont
  {Esposito}}\ and\ \bibinfo {author} {\bibfnamefont {P.}~\bibnamefont
  {Gaspard}},\ }\bibfield  {title} {\bibinfo {title} {Exactly solvable model of
  quantum diffusion},\ }\href {https://doi.org/10.1007/s10955-005-7577-x}
  {\bibfield  {journal} {\bibinfo  {journal} {Journal of Statistical Physics}\
  }\textbf {\bibinfo {volume} {121}},\ \bibinfo {pages} {463} (\bibinfo {year}
  {2005}{\natexlab{a}})}\BibitemShut {NoStop}%
\bibitem [{\citenamefont {Esposito}\ and\ \citenamefont
  {Gaspard}(2005{\natexlab{b}})}]{esposito2005emergence}%
  \BibitemOpen
  \bibfield  {author} {\bibinfo {author} {\bibfnamefont {M.}~\bibnamefont
  {Esposito}}\ and\ \bibinfo {author} {\bibfnamefont {P.}~\bibnamefont
  {Gaspard}},\ }\bibfield  {title} {\bibinfo {title} {Emergence of diffusion in
  finite quantum systems},\ }\href {https://doi.org/10.1103/PhysRevB.71.214302}
  {\bibfield  {journal} {\bibinfo  {journal} {Phys. Rev. B}\ }\textbf {\bibinfo
  {volume} {71}},\ \bibinfo {pages} {214302} (\bibinfo {year}
  {2005}{\natexlab{b}})}\BibitemShut {NoStop}%
\bibitem [{\citenamefont {Eisler}(2011)}]{eisler2011crossover}%
  \BibitemOpen
  \bibfield  {author} {\bibinfo {author} {\bibfnamefont {V.}~\bibnamefont
  {Eisler}},\ }\bibfield  {title} {\bibinfo {title} {Crossover between
  ballistic and diffusive transport: the quantum exclusion process},\ }\href
  {https://doi.org/10.1088/1742-5468/2011/06/p06007} {\bibfield  {journal}
  {\bibinfo  {journal} {Journal of Statistical Mechanics: Theory and
  Experiment}\ }\textbf {\bibinfo {volume} {2011}},\ \bibinfo {pages} {P06007}
  (\bibinfo {year} {2011})}\BibitemShut {NoStop}%
\bibitem [{\citenamefont
  {{\v{Z}}nidari{\v{c}}}(2010{\natexlab{a}})}]{znidaric2010exact}%
  \BibitemOpen
  \bibfield  {author} {\bibinfo {author} {\bibfnamefont {M.}~\bibnamefont
  {{\v{Z}}nidari{\v{c}}}},\ }\bibfield  {title} {\bibinfo {title} {Exact
  solution for a diffusive nonequilibrium steady state of an open quantum
  chain},\ }\href {https://doi.org/10.1088/1742-5468/2010/05/l05002} {\bibfield
   {journal} {\bibinfo  {journal} {Journal of Statistical Mechanics: Theory and
  Experiment}\ }\textbf {\bibinfo {volume} {2010}},\ \bibinfo {pages} {L05002}
  (\bibinfo {year} {2010}{\natexlab{a}})}\BibitemShut {NoStop}%
\bibitem [{\citenamefont {\ifmmode \check{Z}\else
  \v{Z}\fi{}nidari\ifmmode~\check{c}\else \v{c}\fi{}}\ \emph
  {et~al.}(2011)\citenamefont {\ifmmode \check{Z}\else
  \v{Z}\fi{}nidari\ifmmode~\check{c}\else \v{c}\fi{}}, \citenamefont {\ifmmode
  \check{Z}\else \v{Z}\fi{}unkovi\ifmmode~\check{c}\else \v{c}\fi{}},\ and\
  \citenamefont {Prosen}}]{znidaric2011transport}%
  \BibitemOpen
  \bibfield  {author} {\bibinfo {author} {\bibfnamefont {M.}~\bibnamefont
  {\ifmmode \check{Z}\else \v{Z}\fi{}nidari\ifmmode~\check{c}\else
  \v{c}\fi{}}}, \bibinfo {author} {\bibfnamefont {B.}~\bibnamefont {\ifmmode
  \check{Z}\else \v{Z}\fi{}unkovi\ifmmode~\check{c}\else \v{c}\fi{}}},\ and\
  \bibinfo {author} {\bibfnamefont {T.~c.~v.}\ \bibnamefont {Prosen}},\
  }\bibfield  {title} {\bibinfo {title} {Transport properties of a
  boundary-driven one-dimensional gas of spinless fermions},\ }\href
  {https://doi.org/10.1103/PhysRevE.84.051115} {\bibfield  {journal} {\bibinfo
  {journal} {Phys. Rev. E}\ }\textbf {\bibinfo {volume} {84}},\ \bibinfo
  {pages} {051115} (\bibinfo {year} {2011})}\BibitemShut {NoStop}%
\bibitem [{\citenamefont
  {{\v{Z}}nidari{\v{c}}}(2010{\natexlab{b}})}]{znidaric2010a}%
  \BibitemOpen
  \bibfield  {author} {\bibinfo {author} {\bibfnamefont {M.}~\bibnamefont
  {{\v{Z}}nidari{\v{c}}}},\ }\bibfield  {title} {\bibinfo {title} {A matrix
  product solution for a nonequilibrium steady state of an {XX} chain},\ }\href
  {https://doi.org/10.1088/1751-8113/43/41/415004} {\ \textbf {\bibinfo
  {volume} {43}},\ \bibinfo {pages} {415004} (\bibinfo {year}
  {2010}{\natexlab{b}})}\BibitemShut {NoStop}%
\bibitem [{\citenamefont {\ifmmode \check{Z}\else
  \v{Z}\fi{}nidari\ifmmode~\check{c}\else
  \v{c}\fi{}}(2011)}]{znidaric2011solvable}%
  \BibitemOpen
  \bibfield  {author} {\bibinfo {author} {\bibfnamefont {M.}~\bibnamefont
  {\ifmmode \check{Z}\else \v{Z}\fi{}nidari\ifmmode~\check{c}\else
  \v{c}\fi{}}},\ }\bibfield  {title} {\bibinfo {title} {Solvable quantum
  nonequilibrium model exhibiting a phase transition and a matrix product
  representation},\ }\href {https://doi.org/10.1103/PhysRevE.83.011108}
  {\bibfield  {journal} {\bibinfo  {journal} {Phys. Rev. E}\ }\textbf {\bibinfo
  {volume} {83}},\ \bibinfo {pages} {011108} (\bibinfo {year}
  {2011})}\BibitemShut {NoStop}%
\bibitem [{\citenamefont {{\v Z}nidari{\v c}}\ and\ \citenamefont
  {Horvat}(2013)}]{znidaric2013transport}%
  \BibitemOpen
  \bibfield  {author} {\bibinfo {author} {\bibfnamefont {M.}~\bibnamefont {{\v
  Z}nidari{\v c}}}\ and\ \bibinfo {author} {\bibfnamefont {M.}~\bibnamefont
  {Horvat}},\ }\bibfield  {title} {\bibinfo {title} {Transport in a disordered
  tight-binding chain with dephasing},\ }\href
  {https://doi.org/10.1140/epjb/e2012-30730-9} {\bibfield  {journal} {\bibinfo
  {journal} {The European Physical Journal B}\ }\textbf {\bibinfo {volume}
  {86}},\ \bibinfo {pages} {67} (\bibinfo {year} {2013})}\BibitemShut {NoStop}%
\bibitem [{\citenamefont {\ifmmode \check{Z}\else
  \v{Z}\fi{}nidari\ifmmode~\check{c}\else \v{c}\fi{}}\ \emph
  {et~al.}(2016)\citenamefont {\ifmmode \check{Z}\else
  \v{Z}\fi{}nidari\ifmmode~\check{c}\else \v{c}\fi{}}, \citenamefont
  {Scardicchio},\ and\ \citenamefont {Varma}}]{znidaric2016diffusive}%
  \BibitemOpen
  \bibfield  {author} {\bibinfo {author} {\bibfnamefont {M.}~\bibnamefont
  {\ifmmode \check{Z}\else \v{Z}\fi{}nidari\ifmmode~\check{c}\else
  \v{c}\fi{}}}, \bibinfo {author} {\bibfnamefont {A.}~\bibnamefont
  {Scardicchio}},\ and\ \bibinfo {author} {\bibfnamefont {V.~K.}\ \bibnamefont
  {Varma}},\ }\bibfield  {title} {\bibinfo {title} {Diffusive and subdiffusive
  spin transport in the ergodic phase of a many-body localizable system},\
  }\href {https://doi.org/10.1103/PhysRevLett.117.040601} {\bibfield  {journal}
  {\bibinfo  {journal} {Phys. Rev. Lett.}\ }\textbf {\bibinfo {volume} {117}},\
  \bibinfo {pages} {040601} (\bibinfo {year} {2016})}\BibitemShut {NoStop}%
\bibitem [{\citenamefont {Medvedyeva}\ \emph {et~al.}(2016)\citenamefont
  {Medvedyeva}, \citenamefont {Essler},\ and\ \citenamefont
  {Prosen}}]{medvedyeva2016exact}%
  \BibitemOpen
  \bibfield  {author} {\bibinfo {author} {\bibfnamefont {M.~V.}\ \bibnamefont
  {Medvedyeva}}, \bibinfo {author} {\bibfnamefont {F.~H.~L.}\ \bibnamefont
  {Essler}},\ and\ \bibinfo {author} {\bibfnamefont {T.~c.~v.}\ \bibnamefont
  {Prosen}},\ }\bibfield  {title} {\bibinfo {title} {Exact bethe ansatz
  spectrum of a tight-binding chain with dephasing noise},\ }\href
  {https://doi.org/10.1103/PhysRevLett.117.137202} {\bibfield  {journal}
  {\bibinfo  {journal} {Phys. Rev. Lett.}\ }\textbf {\bibinfo {volume} {117}},\
  \bibinfo {pages} {137202} (\bibinfo {year} {2016})}\BibitemShut {NoStop}%
\bibitem [{\citenamefont {Ziolkowska}\ and\ \citenamefont
  {Essler}(2020)}]{ziolkowska2020yangbaxter}%
  \BibitemOpen
  \bibfield  {author} {\bibinfo {author} {\bibfnamefont {A.~A.}\ \bibnamefont
  {Ziolkowska}}\ and\ \bibinfo {author} {\bibfnamefont {F.~H.}\ \bibnamefont
  {Essler}},\ }\bibfield  {title} {\bibinfo {title} {{Yang-Baxter integrable
  Lindblad equations}},\ }\href {https://doi.org/10.21468/SciPostPhys.8.3.044}
  {\bibfield  {journal} {\bibinfo  {journal} {SciPost Phys.}\ }\textbf
  {\bibinfo {volume} {8}},\ \bibinfo {pages} {44} (\bibinfo {year}
  {2020})}\BibitemShut {NoStop}%
\bibitem [{\citenamefont {Essler}\ and\ \citenamefont
  {Piroli}(2020)}]{essler2020integrability}%
  \BibitemOpen
  \bibfield  {author} {\bibinfo {author} {\bibfnamefont {F.~H.~L.}\
  \bibnamefont {Essler}}\ and\ \bibinfo {author} {\bibfnamefont
  {L.}~\bibnamefont {Piroli}},\ }\bibfield  {title} {\bibinfo {title}
  {Integrability of one-dimensional lindbladians from operator-space
  fragmentation},\ }\href {https://doi.org/10.1103/PhysRevE.102.062210}
  {\bibfield  {journal} {\bibinfo  {journal} {Phys. Rev. E}\ }\textbf {\bibinfo
  {volume} {102}},\ \bibinfo {pages} {062210} (\bibinfo {year}
  {2020})}\BibitemShut {NoStop}%
\bibitem [{\citenamefont {Buca}\ \emph {et~al.}(2020)\citenamefont {Buca},
  \citenamefont {Booker}, \citenamefont {Medenjak},\ and\ \citenamefont
  {Jaksch}}]{buca2020dissipative}%
  \BibitemOpen
  \bibfield  {author} {\bibinfo {author} {\bibfnamefont {B.}~\bibnamefont
  {Buca}}, \bibinfo {author} {\bibfnamefont {C.}~\bibnamefont {Booker}},
  \bibinfo {author} {\bibfnamefont {M.}~\bibnamefont {Medenjak}},\ and\
  \bibinfo {author} {\bibfnamefont {D.}~\bibnamefont {Jaksch}},\ }\href@noop {}
  {\bibinfo {title} {Dissipative bethe ansatz: Exact solutions of quantum
  many-body dynamics under loss}} (\bibinfo {year} {2020}),\ \Eprint
  {https://arxiv.org/abs/2004.05955} {arXiv:2004.05955 [cond-mat.stat-mech]}
  \BibitemShut {NoStop}%
\bibitem [{\citenamefont {Bastianello}\ \emph {et~al.}(2020)\citenamefont
  {Bastianello}, \citenamefont {De~Nardis},\ and\ \citenamefont
  {De~Luca}}]{bastianello2020generalized}%
  \BibitemOpen
  \bibfield  {author} {\bibinfo {author} {\bibfnamefont {A.}~\bibnamefont
  {Bastianello}}, \bibinfo {author} {\bibfnamefont {J.}~\bibnamefont
  {De~Nardis}},\ and\ \bibinfo {author} {\bibfnamefont {A.}~\bibnamefont
  {De~Luca}},\ }\bibfield  {title} {\bibinfo {title} {Generalized hydrodynamics
  with dephasing noise},\ }\href {https://doi.org/10.1103/PhysRevB.102.161110}
  {\bibfield  {journal} {\bibinfo  {journal} {Phys. Rev. B}\ }\textbf {\bibinfo
  {volume} {102}},\ \bibinfo {pages} {161110} (\bibinfo {year}
  {2020})}\BibitemShut {NoStop}%
\bibitem [{\citenamefont {Dolgirev}\ \emph {et~al.}(2020)\citenamefont
  {Dolgirev}, \citenamefont {Marino}, \citenamefont {Sels},\ and\ \citenamefont
  {Demler}}]{dolgirev2020nongaussian}%
  \BibitemOpen
  \bibfield  {author} {\bibinfo {author} {\bibfnamefont {P.~E.}\ \bibnamefont
  {Dolgirev}}, \bibinfo {author} {\bibfnamefont {J.}~\bibnamefont {Marino}},
  \bibinfo {author} {\bibfnamefont {D.}~\bibnamefont {Sels}},\ and\ \bibinfo
  {author} {\bibfnamefont {E.}~\bibnamefont {Demler}},\ }\bibfield  {title}
  {\bibinfo {title} {Non-gaussian correlations imprinted by local dephasing in
  fermionic wires},\ }\href {https://doi.org/10.1103/PhysRevB.102.100301}
  {\bibfield  {journal} {\bibinfo  {journal} {Phys. Rev. B}\ }\textbf {\bibinfo
  {volume} {102}},\ \bibinfo {pages} {100301} (\bibinfo {year}
  {2020})}\BibitemShut {NoStop}%
\bibitem [{\citenamefont {Alba}\ and\ \citenamefont
  {Carollo}(2021)}]{alba2021spreading}%
  \BibitemOpen
  \bibfield  {author} {\bibinfo {author} {\bibfnamefont {V.}~\bibnamefont
  {Alba}}\ and\ \bibinfo {author} {\bibfnamefont {F.}~\bibnamefont {Carollo}},\
  }\bibfield  {title} {\bibinfo {title} {Spreading of correlations in markovian
  open quantum systems},\ }\href {https://doi.org/10.1103/PhysRevB.103.L020302}
  {\bibfield  {journal} {\bibinfo  {journal} {Phys. Rev. B}\ }\textbf {\bibinfo
  {volume} {103}},\ \bibinfo {pages} {L020302} (\bibinfo {year}
  {2021})}\BibitemShut {NoStop}%
\bibitem [{\citenamefont {Varma}\ \emph {et~al.}(2017)\citenamefont {Varma},
  \citenamefont {de~Mulatier},\ and\ \citenamefont {\ifmmode \check{Z}\else
  \v{Z}\fi{}nidari\ifmmode~\check{c}\else \v{c}\fi{}}}]{varma2017fractality}%
  \BibitemOpen
  \bibfield  {author} {\bibinfo {author} {\bibfnamefont {V.~K.}\ \bibnamefont
  {Varma}}, \bibinfo {author} {\bibfnamefont {C.}~\bibnamefont {de~Mulatier}},\
  and\ \bibinfo {author} {\bibfnamefont {M.}~\bibnamefont {\ifmmode
  \check{Z}\else \v{Z}\fi{}nidari\ifmmode~\check{c}\else \v{c}\fi{}}},\
  }\bibfield  {title} {\bibinfo {title} {Fractality in nonequilibrium steady
  states of quasiperiodic systems},\ }\href
  {https://doi.org/10.1103/PhysRevE.96.032130} {\bibfield  {journal} {\bibinfo
  {journal} {Phys. Rev. E}\ }\textbf {\bibinfo {volume} {96}},\ \bibinfo
  {pages} {032130} (\bibinfo {year} {2017})}\BibitemShut {NoStop}%
\bibitem [{\citenamefont {Taylor}\ and\ \citenamefont
  {Scardicchio}(2021)}]{taylor2021subdiffusion}%
  \BibitemOpen
  \bibfield  {author} {\bibinfo {author} {\bibfnamefont {S.~R.}\ \bibnamefont
  {Taylor}}\ and\ \bibinfo {author} {\bibfnamefont {A.}~\bibnamefont
  {Scardicchio}},\ }\bibfield  {title} {\bibinfo {title} {Subdiffusion in a
  one-dimensional anderson insulator with random dephasing: Finite-size
  scaling, griffiths effects, and possible implications for many-body
  localization},\ }\href {https://doi.org/10.1103/PhysRevB.103.184202}
  {\bibfield  {journal} {\bibinfo  {journal} {Phys. Rev. B}\ }\textbf {\bibinfo
  {volume} {103}},\ \bibinfo {pages} {184202} (\bibinfo {year}
  {2021})}\BibitemShut {NoStop}%
\bibitem [{\citenamefont {Dorda}\ \emph {et~al.}(2014)\citenamefont {Dorda},
  \citenamefont {Nuss}, \citenamefont {von~der Linden},\ and\ \citenamefont
  {Arrigoni}}]{dorda2014auxiliary}%
  \BibitemOpen
  \bibfield  {author} {\bibinfo {author} {\bibfnamefont {A.}~\bibnamefont
  {Dorda}}, \bibinfo {author} {\bibfnamefont {M.}~\bibnamefont {Nuss}},
  \bibinfo {author} {\bibfnamefont {W.}~\bibnamefont {von~der Linden}},\ and\
  \bibinfo {author} {\bibfnamefont {E.}~\bibnamefont {Arrigoni}},\ }\bibfield
  {title} {\bibinfo {title} {Auxiliary master equation approach to
  nonequilibrium correlated impurities},\ }\href
  {https://doi.org/10.1103/PhysRevB.89.165105} {\bibfield  {journal} {\bibinfo
  {journal} {Phys. Rev. B}\ }\textbf {\bibinfo {volume} {89}},\ \bibinfo
  {pages} {165105} (\bibinfo {year} {2014})}\BibitemShut {NoStop}%
\bibitem [{\citenamefont {Scarlatella}\ \emph {et~al.}(2019)\citenamefont
  {Scarlatella}, \citenamefont {Clerk},\ and\ \citenamefont
  {Schiro}}]{scarlatella2019spectral}%
  \BibitemOpen
  \bibfield  {author} {\bibinfo {author} {\bibfnamefont {O.}~\bibnamefont
  {Scarlatella}}, \bibinfo {author} {\bibfnamefont {A.~A.}\ \bibnamefont
  {Clerk}},\ and\ \bibinfo {author} {\bibfnamefont {M.}~\bibnamefont
  {Schiro}},\ }\bibfield  {title} {\bibinfo {title} {Spectral functions and
  negative density of states of a driven-dissipative nonlinear quantum
  resonator},\ }\href {https://doi.org/10.1088/1367-2630/ab0ce9} {\bibfield
  {journal} {\bibinfo  {journal} {New Journal of Physics}\ }\textbf {\bibinfo
  {volume} {21}},\ \bibinfo {pages} {043040} (\bibinfo {year}
  {2019})}\BibitemShut {NoStop}%
\bibitem [{\citenamefont {Schwarz}\ \emph {et~al.}(2016)\citenamefont
  {Schwarz}, \citenamefont {Goldstein}, \citenamefont {Dorda}, \citenamefont
  {Arrigoni}, \citenamefont {Weichselbaum},\ and\ \citenamefont {von
  Delft}}]{schwarz2016lindblad}%
  \BibitemOpen
  \bibfield  {author} {\bibinfo {author} {\bibfnamefont {F.}~\bibnamefont
  {Schwarz}}, \bibinfo {author} {\bibfnamefont {M.}~\bibnamefont {Goldstein}},
  \bibinfo {author} {\bibfnamefont {A.}~\bibnamefont {Dorda}}, \bibinfo
  {author} {\bibfnamefont {E.}~\bibnamefont {Arrigoni}}, \bibinfo {author}
  {\bibfnamefont {A.}~\bibnamefont {Weichselbaum}},\ and\ \bibinfo {author}
  {\bibfnamefont {J.}~\bibnamefont {von Delft}},\ }\bibfield  {title} {\bibinfo
  {title} {Lindblad-driven discretized leads for nonequilibrium steady-state
  transport in quantum impurity models: Recovering the continuum limit},\
  }\href {https://doi.org/10.1103/PhysRevB.94.155142} {\bibfield  {journal}
  {\bibinfo  {journal} {Phys. Rev. B}\ }\textbf {\bibinfo {volume} {94}},\
  \bibinfo {pages} {155142} (\bibinfo {year} {2016})}\BibitemShut {NoStop}%
\bibitem [{\citenamefont {Citro}\ and\ \citenamefont
  {Mancini}(2018)}]{citro2018out}%
  \BibitemOpen
  \bibfield  {author} {\bibinfo {author} {\bibfnamefont {R.}~\bibnamefont
  {Citro}}\ and\ \bibinfo {author} {\bibfnamefont {F.}~\bibnamefont
  {Mancini}},\ }\href {https://books.google.fr/books?id=xpdmDwAAQBAJ} {\emph
  {\bibinfo {title} {Out-of-Equilibrium Physics of Correlated Electron
  Systems}}},\ Springer Series in Solid-State Sciences\ (\bibinfo  {publisher}
  {Springer International Publishing},\ \bibinfo {year} {2018})\BibitemShut
  {NoStop}%
\bibitem [{\citenamefont {Yueh}\ and\ \citenamefont
  {Cheng}(2006)}]{yueh2006explicit}%
  \BibitemOpen
  \bibfield  {author} {\bibinfo {author} {\bibfnamefont {W.-C.}\ \bibnamefont
  {Yueh}}\ and\ \bibinfo {author} {\bibfnamefont {S.~S.}\ \bibnamefont
  {Cheng}},\ }\bibfield  {title} {\bibinfo {title} {Explicit eigenvalues and
  inverses of several toeplitz matrices},\ }\href
  {https://doi.org/10.1017/S1446181100003424} {\bibfield  {journal} {\bibinfo
  {journal} {The ANZIAM Journal}\ }\textbf {\bibinfo {volume} {48}},\ \bibinfo
  {pages} {73} (\bibinfo {year} {2006})}\BibitemShut {NoStop}%
\bibitem [{\citenamefont {Yueh}\ and\ \citenamefont
  {Cheng}(2008)}]{yueh2008explicit}%
  \BibitemOpen
  \bibfield  {author} {\bibinfo {author} {\bibfnamefont {W.-C.}\ \bibnamefont
  {Yueh}}\ and\ \bibinfo {author} {\bibfnamefont {S.~S.}\ \bibnamefont
  {Cheng}},\ }\bibfield  {title} {\bibinfo {title} {Explicit eigenvalues and
  inverses of tridiagonal toeplitz matrices with four perturbed corners},\
  }\href {https://doi.org/10.1017/S1446181108000102} {\bibfield  {journal}
  {\bibinfo  {journal} {The ANZIAM Journal}\ }\textbf {\bibinfo {volume}
  {49}},\ \bibinfo {pages} {361} (\bibinfo {year} {2008})}\BibitemShut
  {NoStop}%
\bibitem [{\citenamefont {Jin}\ \emph {et~al.}(2020)\citenamefont {Jin},
  \citenamefont {Filippone},\ and\ \citenamefont {Giamarchi}}]{jin2020generic}%
  \BibitemOpen
  \bibfield  {author} {\bibinfo {author} {\bibfnamefont {T.}~\bibnamefont
  {Jin}}, \bibinfo {author} {\bibfnamefont {M.}~\bibnamefont {Filippone}},\
  and\ \bibinfo {author} {\bibfnamefont {T.}~\bibnamefont {Giamarchi}},\
  }\bibfield  {title} {\bibinfo {title} {Generic transport formula for a system
  driven by markovian reservoirs},\ }\href
  {https://doi.org/10.1103/PhysRevB.102.205131} {\bibfield  {journal} {\bibinfo
   {journal} {Phys. Rev. B}\ }\textbf {\bibinfo {volume} {102}},\ \bibinfo
  {pages} {205131} (\bibinfo {year} {2020})}\BibitemShut {NoStop}%
\bibitem [{\citenamefont {Meir}\ and\ \citenamefont
  {Wingreen}(1992)}]{meir1992landauer}%
  \BibitemOpen
  \bibfield  {author} {\bibinfo {author} {\bibfnamefont {Y.}~\bibnamefont
  {Meir}}\ and\ \bibinfo {author} {\bibfnamefont {N.~S.}\ \bibnamefont
  {Wingreen}},\ }\bibfield  {title} {\bibinfo {title} {Landauer formula for the
  current through an interacting electron region},\ }\href
  {https://doi.org/10.1103/PhysRevLett.68.2512} {\bibfield  {journal} {\bibinfo
   {journal} {Phys. Rev. Lett.}\ }\textbf {\bibinfo {volume} {68}},\ \bibinfo
  {pages} {2512} (\bibinfo {year} {1992})}\BibitemShut {NoStop}%
\bibitem [{\citenamefont {Kamenev}(2011)}]{kamenev2011nonequilibrium}%
  \BibitemOpen
  \bibfield  {author} {\bibinfo {author} {\bibfnamefont {A.}~\bibnamefont
  {Kamenev}},\ }\href {https://doi.org/10.1017/CBO9781139003667} {\emph
  {\bibinfo {title} {Field Theory of Non-Equilibrium Systems}}}\ (\bibinfo
  {publisher} {Cambridge University Press},\ \bibinfo {year}
  {2011})\BibitemShut {NoStop}%
\bibitem [{\citenamefont {{T. Jin \textit{et al.}}}(2021)}]{jin2021inprep}%
  \BibitemOpen
  \bibfield  {author} {\bibinfo {author} {\bibnamefont {{T. Jin \textit{et
  al.}}}},\ }\href@noop {} {\bibinfo {title} {to appear}} (\bibinfo {year}
  {2021})\BibitemShut {NoStop}%
\bibitem [{\citenamefont {Scarlatella}\ and\ \citenamefont
  {Schir\'o}(2021)}]{scarlatella2021inprep}%
  \BibitemOpen
  \bibfield  {author} {\bibinfo {author} {\bibfnamefont {O.}~\bibnamefont
  {Scarlatella}}\ and\ \bibinfo {author} {\bibfnamefont {M.}~\bibnamefont
  {Schir\'o}},\ }\href@noop {} {\bibinfo {title} {to be submitted}} (\bibinfo
  {year} {2021})\BibitemShut {NoStop}%
\bibitem [{\citenamefont {Mitra}\ \emph {et~al.}(2006)\citenamefont {Mitra},
  \citenamefont {Takei}, \citenamefont {Kim},\ and\ \citenamefont
  {Millis}}]{MitraEtAlPRL06}%
  \BibitemOpen
  \bibfield  {author} {\bibinfo {author} {\bibfnamefont {A.}~\bibnamefont
  {Mitra}}, \bibinfo {author} {\bibfnamefont {S.}~\bibnamefont {Takei}},
  \bibinfo {author} {\bibfnamefont {Y.~B.}\ \bibnamefont {Kim}},\ and\ \bibinfo
  {author} {\bibfnamefont {A.~J.}\ \bibnamefont {Millis}},\ }\bibfield  {title}
  {\bibinfo {title} {Nonequilibrium quantum criticality in open electronic
  systems},\ }\href {https://doi.org/10.1103/PhysRevLett.97.236808} {\bibfield
  {journal} {\bibinfo  {journal} {Phys. Rev. Lett.}\ }\textbf {\bibinfo
  {volume} {97}},\ \bibinfo {pages} {236808} (\bibinfo {year}
  {2006})}\BibitemShut {NoStop}%
\bibitem [{\citenamefont {Clerk}\ \emph {et~al.}(2010)\citenamefont {Clerk},
  \citenamefont {Devoret}, \citenamefont {Girvin}, \citenamefont {Marquardt},\
  and\ \citenamefont {Schoelkopf}}]{ClerkRMP2010}%
  \BibitemOpen
  \bibfield  {author} {\bibinfo {author} {\bibfnamefont {A.~A.}\ \bibnamefont
  {Clerk}}, \bibinfo {author} {\bibfnamefont {M.~H.}\ \bibnamefont {Devoret}},
  \bibinfo {author} {\bibfnamefont {S.~M.}\ \bibnamefont {Girvin}}, \bibinfo
  {author} {\bibfnamefont {F.}~\bibnamefont {Marquardt}},\ and\ \bibinfo
  {author} {\bibfnamefont {R.~J.}\ \bibnamefont {Schoelkopf}},\ }\bibfield
  {title} {\bibinfo {title} {Introduction to quantum noise, measurement, and
  amplification},\ }\href {https://doi.org/10.1103/RevModPhys.82.1155}
  {\bibfield  {journal} {\bibinfo  {journal} {Reviews of Modern Physics}\
  }\textbf {\bibinfo {volume} {82}},\ \bibinfo {pages} {1155} (\bibinfo {year}
  {2010})}\BibitemShut {NoStop}%
\bibitem [{\citenamefont {Foini}\ \emph {et~al.}(2011)\citenamefont {Foini},
  \citenamefont {Cugliandolo},\ and\ \citenamefont
  {Gambassi}}]{FoiniCugliandoloGambassi_PRB11}%
  \BibitemOpen
  \bibfield  {author} {\bibinfo {author} {\bibfnamefont {L.}~\bibnamefont
  {Foini}}, \bibinfo {author} {\bibfnamefont {L.~F.}\ \bibnamefont
  {Cugliandolo}},\ and\ \bibinfo {author} {\bibfnamefont {A.}~\bibnamefont
  {Gambassi}},\ }\bibfield  {title} {\bibinfo {title} {Fluctuation-dissipation
  relations and critical quenches in the transverse field ising chain},\ }\href
  {https://doi.org/10.1103/PhysRevB.84.212404} {\bibfield  {journal} {\bibinfo
  {journal} {Phys. Rev. B}\ }\textbf {\bibinfo {volume} {84}},\ \bibinfo
  {pages} {212404} (\bibinfo {year} {2011})}\BibitemShut {NoStop}%
\bibitem [{\citenamefont {Dalla~Torre}\ \emph {et~al.}(2012)\citenamefont
  {Dalla~Torre}, \citenamefont {Demler}, \citenamefont {Giamarchi},\ and\
  \citenamefont {Altman}}]{DallaTorreetalPRB2012}%
  \BibitemOpen
  \bibfield  {author} {\bibinfo {author} {\bibfnamefont {E.~G.}\ \bibnamefont
  {Dalla~Torre}}, \bibinfo {author} {\bibfnamefont {E.}~\bibnamefont {Demler}},
  \bibinfo {author} {\bibfnamefont {T.}~\bibnamefont {Giamarchi}},\ and\
  \bibinfo {author} {\bibfnamefont {E.}~\bibnamefont {Altman}},\ }\bibfield
  {title} {\bibinfo {title} {Dynamics and universality in noise-driven
  dissipative systems},\ }\href {https://doi.org/10.1103/PhysRevB.85.184302}
  {\bibfield  {journal} {\bibinfo  {journal} {Phys. Rev. B}\ }\textbf {\bibinfo
  {volume} {85}},\ \bibinfo {pages} {184302} (\bibinfo {year}
  {2012})}\BibitemShut {NoStop}%
\bibitem [{\citenamefont {Schir\'o}\ and\ \citenamefont
  {Mitra}(2014)}]{SchiroMitraPRL14}%
  \BibitemOpen
  \bibfield  {author} {\bibinfo {author} {\bibfnamefont {M.}~\bibnamefont
  {Schir\'o}}\ and\ \bibinfo {author} {\bibfnamefont {A.}~\bibnamefont
  {Mitra}},\ }\bibfield  {title} {\bibinfo {title} {Transient orthogonality
  catastrophe in a time-dependent nonequilibrium environment},\ }\href
  {https://doi.org/10.1103/PhysRevLett.112.246401} {\bibfield  {journal}
  {\bibinfo  {journal} {Phys. Rev. Lett.}\ }\textbf {\bibinfo {volume} {112}},\
  \bibinfo {pages} {246401} (\bibinfo {year} {2014})}\BibitemShut {NoStop}%
\bibitem [{\citenamefont {Sieberer}\ \emph {et~al.}(2016)\citenamefont
  {Sieberer}, \citenamefont {Buchhold},\ and\ \citenamefont
  {Diehl}}]{SiebererRepProgPhys2016}%
  \BibitemOpen
  \bibfield  {author} {\bibinfo {author} {\bibfnamefont {L.~M.}\ \bibnamefont
  {Sieberer}}, \bibinfo {author} {\bibfnamefont {M.}~\bibnamefont {Buchhold}},\
  and\ \bibinfo {author} {\bibfnamefont {S.}~\bibnamefont {Diehl}},\ }\bibfield
   {title} {\bibinfo {title} {Keldysh field theory for driven open quantum
  systems},\ }\href {https://doi.org/10.1088/0034-4885/79/9/096001} {\bibfield
  {journal} {\bibinfo  {journal} {Reports on Progress in Physics}\ }\textbf
  {\bibinfo {volume} {79}},\ \bibinfo {pages} {096001} (\bibinfo {year}
  {2016})}\BibitemShut {NoStop}%
\bibitem [{\citenamefont {{X. Turkeshi \textit{et
  al.}}}(2021)}]{turkeshi2021inprep}%
  \BibitemOpen
  \bibfield  {author} {\bibinfo {author} {\bibnamefont {{X. Turkeshi \textit{et
  al.}}}},\ }\href@noop {} {\bibinfo {title} {to be submitted}} (\bibinfo
  {year} {2021})\BibitemShut {NoStop}%
\end{thebibliography}%

\end{document}